\documentclass[
journal=jctcce, 
manuscript=article]{achemso}

\usepackage[version=3]{mhchem} 
\usepackage{amsmath}
\usepackage{amssymb}
\usepackage{natbib}
\usepackage{graphicx}
\usepackage{subcaption}
\usepackage{longtable} 
\usepackage{notes2bib}
\usepackage{url}
\usepackage{array}
\usepackage{xr-hyper}
\usepackage{hyperref}
\usepackage{multirow}
\usepackage{marginnote}
\usepackage{bm}
\usepackage{physics}
\usepackage{braket}
\usepackage{xcolor}
\usepackage{soul}

\makeatletter
\newcommand*{\addFileDependency}[1]{
  \typeout{(#1)}
  \@addtofilelist{#1}
  \IfFileExists{#1}{}{\typeout{No file #1.}}
}
\makeatother

\newcommand*{\myexternaldocument}[1]{%
    \externaldocument{#1}%
    \addFileDependency{#1.tex}%
    \addFileDependency{#1.aux}%
}

\myexternaldocument{si}

\title{Green's function formulation of quantum defect embedding theory}

\author{Nan Sheng}
\affiliation
{Department of Chemistry, University of Chicago, Chicago, IL 60637, USA.}
\altaffiliation{These two authors contributed equally.}

\author{Christian Vorwerk}
\affiliation
{Pritzker School of Molecular Engineering, University of Chicago, Chicago, IL 60637, USA.}
\altaffiliation{These two authors contributed equally.}

\author{Marco Govoni}
\affiliation
{Pritzker School of Molecular Engineering, University of Chicago, Chicago, IL 60637, USA.}
\alsoaffiliation
{Materials Science Division and Center for Molecular Engineering, Argonne National Laboratory, Lemont, IL 60439, USA.}
\email{mgovoni@anl.gov}

\author{Giulia Galli}
\affiliation
{Pritzker School of Molecular Engineering, University of Chicago, Chicago, IL 60637, USA.}
\alsoaffiliation
{Department of Chemistry, University of Chicago, Chicago, IL 60637, USA.}
\alsoaffiliation[MSD]
{Materials Science Division and Center for Molecular Engineering, Argonne National Laboratory, Lemont, IL 60439, USA.}
\email{gagalli@uchicago.edu}

\begin{document}

\begin{abstract}
 We present a Green's function formulation of the quantum defect embedding theory (QDET) where a double counting scheme is rigorously derived within the $G_0W_0$ approximation. We then show the robustness of our methodology by applying the theory with the newly derived scheme to several defects in diamond. Additionally, we discuss a strategy to obtain converged results as a function of the size and composition of the active space. Our results show that QDET is a promising approach to investigate strongly correlated states of defects in solids.
\end{abstract}

\section{Introduction} \label{sec:introduction}
Electronic structure calculations of solids and molecules rely on the solution
of approximate forms of the Schr\"{o}dinger equation, for example using density
functional
theory\cite{hohenberg1964,kohn1965,neugebauer2013,mardirossian2017,jones2015,burke2012},
many-body perturbation theory\cite{strinati1988,martin2016,golze2019},  or
quantum chemistry methods\cite{bartlett2007,zhang2019,helgaker2014} and, in some
cases, quantum Monte Carlo\cite{martin2016}. Employing theoretical
approximations is almost always necessary, as the solution of the electronic
structure problem using the full many-body Hamiltonian of the system is still
prohibitive, from a computational standpoint,  for most molecules and solids,
even in the case of the time independent Schr\"{o}edinger equation.

Interestingly, there are important problems in condensed matter physics,
materials science and chemistry for which a specific region of interest may be
identified, a so-called  active region, surrounded by a host medium, and for
which the electronic structure problem can be solved at a high level of theory,
for example, by exact diagonalization. An active region may be associated, for
instance, to point defects in materials, active site of catalysts or
nanoparticles embedded in soft or hard matrices. All of these problems may then
be addressed using embedding theories
\cite{sun2016,jones2020,sheng_quantum_2021} which separate the electronic
structure problem of the active region from that of the host environment.  Each
part of the system is described at the quantum-mechanical level\cite{sun2016},
at variance with quantum embedding models, e.g. QM/MM, where only the active
space or region is described with quantum-mechanical methods, while the
environment is treated classically~\cite{senn2009,liu2014}.

Several embedding techniques have been proposed in the literature in recent
years, which may be classified by the level of theory chosen to describe the
different portions of the system. Density-based theories encompass
density-functional-theory embedding in-density- functional-theory (DFT-in-DFT)
and wavefunction embedding in-DFT
(WF-in-DFT)~\cite{libisch2014,gomes2008,goodpaster2012}. In these schemes the
environment is described within DFT and the active region with a level of DFT
higher than that adopted for the environment or with quantum-chemical,
wave-function based methods. Density-matrix embedding theory (DMET)
\cite{wouters2016,knizia2012,knizia2013,cui_efficient_2020,pham2020,hermes2019,pham2018} employs
instead the density matrix of the system to define an embedding protocol.
Finally, Green’s ($G$) function-based quantum embedding methods include  the
self-energy embedding\cite{lan2017,zgid2017,rusakov2019} , dynamical mean field
(DMFT) \cite{georges1992,georges1996,georges2004,anisimov1997,kotliar2006} and
the quantum defect embedding theories (QDET)
\cite{ma2020a,ma2021,sheng_quantum_2021}.

QDET is a theory we have recently proposed for the calculation of defect
properties in solids, with the goal of computing strongly correlated states
which may not be accurately obtained using  mean field theories, such as DFT,
when using large supercells. However, the applicability of the theory is not
restricted to defects in solids and QDET may be used to study, in general, a
small guest region embedded in a large host condensed system. Similar to all
Green’s function based methods, in QDET the active space is defined by a set of
single-particle electronic states. The set includes the states localized in
proximity of the defect or impurity and, in some cases, contains additional
single-particle orbitals.

The embedding protocol used in QDET leads to a delicate problem that many
embedding theories have in common, at least conceptually: the presence  of
“double counting” terms in the effective Hamiltonian of the active regions.
These are terms that are computed both at the level of theory chosen for the
active region and at the lower level chosen for the environment. Hence
corrections (often called double counting corrections) are required to restore
the accuracy of the effective Hamiltonian.

In the original formulation of QDET presented in Ref. \citenum{ma2021}, we
adopted an approximate double counting correction based on Hartree-Fock theory.
Here we present a more rigorous derivation of QDET based on Green's functions
and we derive a double counting correction that is exact within the $G_0 W_0$
approximation and when retardation effects are neglected. We call this
correction EDC@$G_0 W_0$ ({\it exact} double counting at $G_0W_0$ level of theory). We then apply QDET with the EDC@$G_0 W_0$
scheme to several spin defects in solids and we present a strategy to
systematically converge the results as a function of the composition and size of
the active space. Finally we show that using the EDC@$G_0 W_0$, we obtain results for the
electronic  structure of spin defects consistent with experiments and in good
agreement with results obtained with other embedding theories \cite{mitra2021}.

The rest of the paper is organized as follows. Section \ref{sec:theory} presents
the formulation of QDET and Section \ref{sec:implementation} its implementation.
Our results are presented in Section \ref{sec:results} and conclusions in
Section \ref{sec:conclusions}.

\section{Formulation of quantum defect embedding theory (QDET)}\label{sec:theory}

In Ref.~\citenum{ma2020a,ma2021} we introduced the Quantum Defect Embedding
Theory (QDET), an embedding scheme that describes a condensed system where the
electronic excitations of interest occur within a small subspace (denoted as the
active space $A$) of the full Hilbert space of the system. The formulation is
based on the description of the system using periodic DFT and assumes that the
interaction between active regions belonging to  periodic replicas may be
neglected (\textit{i.e} the dilute limit).  We summarize below the original formulation of the
QDET method, including the approximate double counting scheme  used in
Ref.~\citenum{ma2020a,ma2021}. We then present a Green's function formulation of
QDET which enables the definition of an improved correction to the double counting scheme originally adopted, which is exact within the $G_0W_0$
approximation.

In QDET an active space is defined as the space spanned by an orthogonal set of
functions $\{\zeta_i\}$, for instance, selected eigenstates of the Kohn-Sham
(KS) Hamiltonian describing a solid, or localized functions, \textit{e.g.},
Wannier orbitals constructed from Kohn-Sham eigenstates through a unitary
transformation. Within the Born-Oppenheimer and nonrelativisitic approximations,
the many-body effective Hamiltonian of a system of interacting electrons within
a given active space takes the following form:
\begin{equation}\label{eq:def_hameff}
  H^{\mathrm{eff}} = \sum_{ij}^A t^{\mathrm{eff}}_{ij}a^\dagger_i a_j +
  \frac{1}{2} \sum_{ijkl}^A v^{\mathrm{eff}}_{ijkl}
  a^\dagger_i a^\dagger_j a_l a_k,
\end{equation}
where $t^{\mathrm{eff}}$ and $v^{\mathrm{eff}}$ are  one- and two-body terms
that include the influence of the environment on the chosen active space.  In
QDET, these terms  are determined by first carrying out a mean-field calculation
of the full solid using, \textit{e.g.}, DFT. Once the KS eigenstates and
eigenvalues of the full system are obtained, the two-body terms
$v^{\mathrm{eff}}$ are computed as the matrix elements of the partially screened
static Coulomb potential $W^R_0$, \textit{i.e.},
\begin{equation}\label{eq:def_veff}
  v^{\mathrm{eff}}_{ijkl} = \left[W^R_0 \right]_{ijkl} := \int d\mathbf{x}d\mathbf{x}'
  \zeta_i(\mathbf{x})\zeta_k(\mathbf{x})W^R_0(\mathbf{x},\mathbf{x}'; \omega=0)
  \zeta_j(\mathbf{x}')\zeta_l(\mathbf{x}').
\end{equation}
The term $W^R_0$ in Eq.~\ref{eq:def_veff} is
obtained by screening the bare Coulomb potential $v$ with the reduced
polarizability $P^{R}_0$,  defined by the following equation:
\begin{equation}\label{eq:def_we}
  W^R_0 = v +v P^{R}_0W^R_0.
\end{equation}
The reduced polarizability may be obtained by subtracting from the total
irreducible polarizability of the periodic system the contribution from the
active space, namely $P^{R}_0 = P_0 - P^A_0$. 
Within  the Random-Phase Approximation (RPA), the active space polarizability
$P^A_0$ is given by
\begin{equation}\label{eq:def_p0a}
  \begin{aligned}
  P^A_0 (\mathbf{x}_1, \mathbf{x}_2; \omega ) = \sum_i^{\mathrm{occ}}
    \sum_j^{\mathrm{unocc}}& \left(f^A \psi_i^{\mathrm{KS}}\right)(\mathbf{x}_1)
  \left(f^A \psi_j^{\mathrm{KS}}\right)(\mathbf{x}_1)
  \left(f^A \psi_j^{\mathrm{KS}}\right)(\mathbf{x}_2)
  \left(f^A \psi_i^{\mathrm{KS}}\right)(\mathbf{x}_2)\\
    &\times\left(
    \frac{1}{\omega-(\epsilon_j^{\mathrm{KS}}-\epsilon_i^{\mathrm{KS}})+\mathrm{i}\eta}
    +\frac{1}{\omega+(\epsilon_j^{\mathrm{KS}}-\epsilon_i^{\mathrm{KS}})-\mathrm{i}\eta}
    \right),
  \end{aligned}
\end{equation}
where $\psi^{\mathrm{KS}}_i$ and $\epsilon^{\mathrm{KS}}_i$ are the Kohn-Sham
eigenfunctions and -values, respectively and ``occ'' and ``unocc'' denote sums
over occupied and empty states, respectively. Here, we have introduced the
projector $f^A = \sum^A_i | \zeta_i \rangle \langle \zeta_i |$ on the active
space. An expression of the total irreducible polarizability may be obtained by
omitting the projectors $f^A$ on the RHS of Eq.~\ref{eq:def_p0a}. In
Refs.~\citenum{ma2020a,ma2021} we proposed an efficient implementation of
Eq.~\ref{eq:def_p0a} that does not require any explicit summation over
unoccupied states, thus enabling the application of QDET  to large systems.

The definition of $v^{\mathrm{eff}}$ given above includes contributions to the
Hartree and exchange correlation energies that are also included in the DFT
calculations for the whole solid, \textit{ i.e.}, it contains  so-called
double counting (dc) terms.  The latter are subtracted (that is corrections to
double counting contributions are applied) when defining the  one-body terms
$t^{\mathrm{eff}}$, 
\begin{equation}\label{eq:intro_teff}
  t^{\mathrm{eff}}_{ij} = H^{\mathrm{KS}}_{ij} - t^{\mathrm{dc}}_{ij},
\end{equation}
where $H^{\mathrm{KS}}$ is the Kohn-Sham Hamiltonian. 

\subsection{QDET based on density functional theory} 
In previous applications of QDET, the double counting term $t^{\mathrm{dc}}$ was
approximated, since within a DFT formulation of the theory, one
cannot define an explicit expression for the exchange and correlation potential
for a subset of electronic states.  Therefore, an approximate form of
$t^{\mathrm{dc}}$ inspired by Hartree-Fock was used, given by:
\begin{equation}
  t^{\mathrm{dc}}_{ij} \approx \sum_{kl}^A \left( \left[ W^R_0 (\omega = 0) \right]_{ikjl}  - 
  \left[W^R_0 (\omega = 0) \right]_{ijkl} \right) \rho^A_{kl}, \label{eq:t_hf}
\end{equation}
where the reduced density matrix of the active space $A$ is given by
$\rho^A_{ij}=\sum_{k}^{\mathrm{occ}} \langle \zeta_i | \psi_k \rangle \langle
\psi_k | \zeta_j \rangle$.

Once the terms in the Hamiltonian of Eq.~\ref{eq:def_hameff} are defined, the
electronic structure of the correlated states in the active space $A$ is
obtained  from an exact diagonalization procedure,  using the full configuration
interaction (FCI) method.

We note that within Hartree-Fock theory, the expression of $t^{\mathrm{dc}}$,
where  $v$  replaces $W_0^R$ on the RHS of Eq.~\ref{eq:t_hf}, is exact; however
Eq. 6 turns out to be  an approximate expression when used within DFT.  While
QDET with an approximate double counting scheme has been successfully applied to
a range of defects in diamond and SiC, the influence of the approximation used
for $t^{\mathrm{dc}}$ deserves further scrutiny. Most importantly, a formulation
without double counting approximations  is desirable. 

In the next section,  we present a  Green's function formulation of QDET and we
derive  an analytical expression for $t^{\mathrm{eff}}$ that in turns leads to
an expression for the double counting term $t^{\mathrm{dc}}$ which is exact
within the $G_0W_0$ approximation.  

\subsection{Green's function formulation of QDET}
Instead of starting by a DFT formulation,  we describe the  interacting
electrons in a solid by defining the one-body Green's function $G$ and  the
screened Coulomb potential $W$. The reason to introduce a Green's function
description stems from the fact that the self-energy $\Sigma$ and its
irreducible polarizability $P$ can be written as sums of contributions from
different portions of the entire system. The basic equations relating $G$, $W$,
$\Sigma$ and $P$ are:
\begin{align}
  G^{-1} &= g^{-1} - \Sigma \label{eq:g_def} \\
  W^{-1} &= v^{-1} - P \label{eq:w_def},
\end{align}
where $v$ is the bare Coulomb potential; the bare Green's function $g= (\omega -
h)^{-1} $ with  $h = -\frac{1}{2} \nabla^2 + v_{\mathrm{ion}}$, where
$v_{\mathrm{ion}}$ is the electrostatic potential of the nuclei.

We chose two different levels of theory to describe different portions of the
system, namely we describe the active space with a so-called higher level theory
(high) than that applied to the whole system (low).  We write the self-energy
and polarizability of the whole system as:
\begin{align}
  \Sigma &= \Sigma^{\mathrm{low}} + \left( \Sigma^{\mathrm{high}} -
  \Sigma^{\mathrm{dc}} \right)_A \label{eq:sigma_split}\\
  P &= P^{\mathrm{low}} + \left( P^{\mathrm{high}} - P^{\mathrm{dc}}
  \right)_A.\label{eq:p_split}
\end{align}
Here, we introduced the double counting terms $\Sigma^{\mathrm{dc}}$ and
$P^{\mathrm{dc}}$ that correct for the contributions to the self-energy and
polarizability of the active space $A$, which are included both in the high- and
low-level descriptions of  $A$. The terms with subscript $A$ in Eqs.
~\ref{eq:sigma_split} and \ref{eq:p_split} are defined in the subspace $A$.
Inserting Eqs.~\ref{eq:sigma_split} and \ref{eq:p_split} into Eq.~\ref{eq:g_def}
and \ref{eq:w_def}, respectively, leads to
\begin{align}
  G^{-1} &= \left[ G^R \right]^{-1} - \Sigma^{\mathrm{high}}_A \label{eq:g_split} \\
  W^{-1} &= \left[ W^R \right]^{-1} - P^{\mathrm{high}}_A \label{eq:w_split},
\end{align}
where we have defined the renormalized Green's function $G^R$ and partially
screened potential $W^R$ as:
\begin{align}
  \left[ G^R \right]^{-1} &= g^{-1} - \Sigma^{\mathrm{low}} +
  \Sigma^{\mathrm{dc}}_A \\
  \left[ W^R \right]^{-1} &= v^{-1} - P^{\mathrm{low}} + P^{\mathrm{dc}}_A. \label{eq:WRm1} 
\end{align}
Comparing Eq.~\ref{eq:g_split} and \ref{eq:w_split} with Eq.~\ref{eq:g_def} and
\ref{eq:w_def}, we find that the problem of determing $G$ and $W$ for the total
system $A+E$ has been simplified: only a solution with the high-level method
within $A$ is necessary to obtain $G$ and $W$. To obtain such solution, the bare
Green's function $g$ and bare Coulomb potential $v$ should be replaced by their
renormalized counterparts $G^R$ and $W^R$, respectively. We now turn to derive
expressions for $\Sigma^{\mathrm{low}}$ and $P^{\mathrm{low}}$, and
$\Sigma^{\mathrm{high}}$ and $P^{\mathrm{high}}$,  which will then allow for the
definition of all terms entering the  effective Hamiltonian of the active space.

\subsubsection{Effective Hamiltonian} 
Under the assumption that retardation effects may be neglected, \textit{ i.e.},
assuming that one- and two-body interactions within $A$ are instantaneous, we
can  derive a simple equation relating $G^{R}$ and $W^{R}$ and the parameters of
the effective Hamiltonian in Eq.~\ref{eq:def_hameff}.  In the absence of
retardation effects, the effective Green's function $G^R$ is given by the
Lehmann representation and we have:
\begin{equation}\label{eq:def_teff}
  G^R(\omega) \approx \left[ \omega - t^{\mathrm{eff}} \right]^{-1}.
\end{equation}
Note that the validity of this equation rests on the assumption that the
non-diagonal terms of the self-energy coupling the active space and the
environment are negligible, i.e.   $\left(\Sigma^{\mathrm{low}}\right)_{AE} =
0$.  Eq.~\ref{eq:def_teff} defines the one-body terms $t^{\mathrm{eff}}$ of
Eq.~\ref{eq:def_hameff}. To derive  the two-body terms, we neglect the frequency
dependence of the screened Coulomb interaction and write: 
\begin{equation}\label{eq:gen_veff}
  v^{\mathrm{eff}}_{ijkl} \approx \int d\mathbf{x}d\mathbf{x}'
  \zeta_i(\mathbf{x})\zeta_k(\mathbf{x})W^R(\mathbf{x},\mathbf{x}', \omega=0)
  \zeta_j(\mathbf{x}')\zeta_l(\mathbf{x}').
\end{equation}
We note that the static approximation of the screened Coulomb interaction is
commonly employed to calculate neutral excitations in solids and molecules
within many-body perturbation theory (MBPT)\cite{martin2016}, and it has been
shown to yield neutral excitations in a wide range of materials with excellent
accuracy.  In order to obtain the expressions of $t^{\mathrm{eff}}$ and
$v^\mathrm{eff}$ for which an analytic expression of $t^{\mathrm{dc}}$ can be
derived, we turn to the $G_0W_0$ approximation, which we use as low level of
theory for the entire system. Such a choice of  low-level theory enables the
separation of the self-energy as required by Eq.~\ref{eq:sigma_split}.

\subsubsection{$G_0W_0$ approximation as low-level theory}
We use  the $G_0W_0$ approximation and write
$\Sigma^{\mathrm{low}}=\Sigma^{G_0W_0}$ with
\begin{equation}\label{eq:$G_0W_0$}
  \Sigma^{G_0W_0} = V_{\mathrm{H}} + \Sigma_{\mathrm{xc}} = v \rho + \mathrm{i}G_0W_0.
\end{equation}
(See SI for additional details).
We evaluate the Green's function $G_0$  using the KS Hamiltonian, \textit{i.e.}, 
\begin{equation}\label{eq:g0}
    G_0(\omega) = (\omega - H^{\mathrm{KS}})^{-1} \,.
\end{equation}
The screened Coulomb potential is
obtained as $W_0^{-1} = v^{-1} - P_0$, with 
\begin{equation}
  P_0 = -\mathrm{i}G_0G_0. \label{eq:iGG}
\end{equation}

\subsubsection{Double counting correction}\label{sec:dc}
To derive the double counting terms $\Sigma^{\mathrm{dc}}$ and $P^{\mathrm{dc}}$, we require that the chain rule be satisfied, \textit{i.e.}, that when using the low-level of theory to describe both the total system ($A$+$E$) and the active space ($A$),  the total self-energy and the total polarizability on the LHS of Eq. \ref{eq:sigma_split} and \ref{eq:p_split} are the same as $\Sigma^{\mathrm{low}}$ and $P^{\mathrm{low}}$, respectively. This requirement implies that $\Sigma^{\mathrm{dc}}$ and $P^{\mathrm{dc}}$ coincide with the self-energy and the polarizability derived from the effective Hamiltonian expressed at the low-level of theory. Within the $G_0W_0$ approximation, this requirement leads to the following expressions


\begin{align}
& \Sigma^{\mathrm{dc}}  = \Sigma_{G_0W_0}^{\mathrm{eff}} \label{eq:cr1} \\
& P^{\mathrm{dc}} = P_0^{\mathrm{eff}}. \label{eq:cr2}
\end{align}
Here the superscript `$\mathrm{eff}$' indicates that the the self-energy and polarizabilities are computed for $H^{\mathrm{eff}}$; these quantities are different from the corresponding ones evaluated for $H$ projected onto $A$. The double counting contribution to the
polarizability,  $P^{\mathrm{dc}}$, is obtained with Eq.~\ref{eq:iGG} after
restricting the Green's function to $A$, \textit{ i.e.},
\begin{equation}\label{eq:p_dc}
  P^{\mathrm{dc}} = P_0^A = -\mathrm{i}G_0^A G_0^A,
\end{equation}
with $G_0^A = f^A G_0 f^A$. 
Eq.~\ref{eq:WRm1} and~\ref{eq:p_dc} allow us to determine the partially screened
Coulomb potential $W^R_0$ as
\begin{equation}\label{eq:def_we2}
  \left[ W^R_0 \right]^{-1} = 
  v^{-1} - \left( G_0^A G_0^{R} + G_0^{R} G_0^A + G_0^{R}
  G_0^{R}\right) = v^{-1} - P_0^R, 
\end{equation}
where we have defined the reduced Kohn-Sham Green's function $G_0^{R} = G_0-
G_0^A$. The matrix elements of $W_0^R$ thus enter the definition of the two-body
terms of the effective Hamiltonian. Hence we have shown that by framing QDET
within the context of Green's embedding theories we recover Eq.~\ref{eq:def_we}.

Similar to the derivation of $P^{\mathrm{dc}}$, the double counting contribution
to the self-energy, $\Sigma^{\mathrm{dc}}$, is given by the $G_0W_0$ self-energy associated to $H^{\mathrm{eff}}$, \textit{i.e.}, $\Sigma^{\mathrm{eff}}_{G_0W_0}= v^{\mathrm{eff}}\rho^A +\mathrm{i}G_0^AW^{\mathrm{eff}}_0$, where $v^{\mathrm{eff}} = W^R$, $\rho^{\mathrm{eff}} = \rho^A$, $G_0^{\mathrm{eff}} = G_0^A$, and $W_0^{\mathrm{eff}} = W_0$ (the derivation is reported in Sec. 2 of the SI). The final result reads
\begin{equation}\label{eq:sigma_dc}
  \Sigma^{\mathrm{dc}} = W^R_0 \rho^A + \mathrm{i}G_0^A  W_0.
\end{equation}
We note that in the second term of Eq.~\ref{eq:sigma_dc}, the
screened Coulomb potential in $A$ is obtained by adding to $W^R_0$ the screening
potential generated by the polarizability $P_0^A$. This addition yields by
definition the total screened Coulomb potential, since $W^{-1}_0 = \left[ W^R_0
\right]^{-1} - P_0^A$.
We note that the chain rule by construction leads to a Hartree double-counting self-energy (the first term of Eq.~\ref{eq:sigma_dc}) that is defined in terms of the partially screened Coulomb potential. This is essential to remove the Hartree self-energy of the effective Hamiltonian (see Sec.~2 in the SI) that is already accounted for in the $G_0W_0$ calculation of the total system ($A$+$E$). Equivalently, the second term in Eq.~\ref{eq:sigma_dc} removes the exchange-correlation self-energy of the effective Hamiltonian at the $G_0W_0$ level, as this self-energy has already been accounted for in the $G_0W_0$ calculation of the total system.

Having obtained  explicit expressions for the double counting terms, we can
finally determine the one-body parameters of the effective Hamiltonian. We write
$G^R$ as:
\begin{equation}
  \begin{aligned}
    \left[ G^R \right]^{-1} &= g^{-1} - \left[ V_{\mathrm{H}} + \mathrm{i}G_0W_0
    \right] +
    \left[ W^R_0 (\omega = 0) \rho + \mathrm{i}G_0^AW_0 \right]\\ 
    &= \omega - H^{\mathrm{KS}}
    + V_{\mathrm{xc}} + W^R_0 \rho^A - \mathrm{i}G_0^R W_0.
  \end{aligned}
\end{equation}
By comparing the equation above with Eqs.~\ref{eq:intro_teff} and
\ref{eq:def_teff}, we obtain the double counting contribution to the effective
one-body terms as:
\begin{equation}\label{eq:teff_final}
  t^{\mathrm{dc}} = V_{\mathrm{xc}} + W^R_0(\omega = 0) \rho^A 
  - \mathrm{i}G_0^R W_0.
\end{equation}
In general, the one-body terms should be frequency-dependent, due to the
frequency dependency of $G_0^R(\omega)$ and $W_0(\omega)$. To obtain static
expressions for the one-body terms, we evaluate $\mathrm{i}G_0^RW_0$ at the
quasi-particle energies. More details are provided in
Sec.~\ref{sec:implementation}. 

As we will see in Sec.~\ref{sec:results}, the double counting scheme defined
here yields more accurate results that the Hartee-Fock one, since it satisfies
the chain rule by construction. On the contrary, the  Hartree-Fock double
counting scheme used in
Ref.~\citenum{bockstedte_ab_2018,ma2020,ma2020a,ma2021,muechler_quantum_2021,pfaffle_screened_2021}
does not satisfy the chain rule and thus may introduce  errors originating from
the separation between active space and environment.  A derivation of
the Hartree-Fock double counting within the Green's function formalism is
provided in the SI.

\section{Implementation} \label{sec:implementation}
The QDET method and the  double counting scheme of Eq.~\ref{eq:teff_final}  are implemented in
the  WEST\cite{govoni2015} (Without Empty States) code, a massively parallel
open-source code designed for large-scale MBPT calculations of complex
condensed-phase systems, such as defective solids.  In the  WEST code, a
separable form of $W_0$ is obtained using the projected eigen-decomposition of
the dielectric matrix (PDEP) ~\cite{wilson_efficient_2008,govoni2015}, which
avoids the inversion and storage of large dielectric matrices. Importantly,
explicit summations over empty KS orbitals entering the expressions of $P_0$ and
$G_0$ are eliminated using density functional perturbation theory
(DFPT)~\cite{baroni2001} and the Lanczos method~\cite{umari_gw_2010,nguyen2012},
respectively.  The implementation of $W_0^R$ in WEST has been reported
previously~\cite{ma2021}. In the following, we focus on the implementation of
the double counting term entering  Eq.~\ref{eq:teff_final}.

In our current implementation, the active space $A$ is defined by a set of
Kohn-Sham eigenstates, and  $G^R_0$ is given  by
\begin{equation}
  G_0^R(\mathbf{x},\mathbf{x}';\omega) = \sum_i^{E} \frac{\psi^{\mathrm{KS}}_i(\mathbf{x})
  \psi^{\mathrm{KS}}_i(\mathbf{x}')}{\omega - \epsilon^{\mathrm{KS}}_i +
  \mathrm{i}\eta \; \mathrm{sgn}(\epsilon^{\mathrm{KS}}_i- \epsilon_\mathrm{F})},
\end{equation}
where $E$ is the environment, $\mathrm{sgn}$ is the sign function and $\epsilon_\mathrm{F}$ is the Fermi
energy.
The term $\Delta \Sigma_{\mathrm{xc}} =
\mathrm{i}G^R_0W_0$ in Eq.~\ref{eq:teff_final} is  given by
\begin{equation}\label{eq:freq_int}
  \Delta \Sigma_{\mathrm{xc}}(\omega) = \mathrm{i} \int \frac{d \omega'}{2 \pi}
  G^R_0(\omega+\omega')W_0(\omega'),
\end{equation}
where the integration is performed using a contour deformation
technique\cite{godby1988,giantomassi2011,govoni2015}. Finally, to obtain static
double counting terms, we evaluate Eq.~\ref{eq:freq_int} at the quasiparticle
energies
$\epsilon^{\mathrm{QP}}_i$, \textit{ i.e.},
\begin{equation}
  \left[ \Delta \Sigma_{\mathrm{xc}} \right]_{ij} = \frac{1}{2} \mathrm{Re} \; \left[
    \left[ \Delta \Sigma_{\mathrm{xc}} \right]_{ij} (\epsilon^{\mathrm{QP}}_i) + 
    \left[ \Delta \Sigma_{\mathrm{xc}} \right]_{ij}(\epsilon^{\mathrm{QP}}_j) \right],
    \label{eq:staticsigma}
\end{equation}
where the quasi-particle energies are obtained by solving iteratively the
equation 
$\epsilon^{\mathrm{QP}}_i = \epsilon^{\mathrm{KS}}_i + \langle
  \psi^{\mathrm{KS}}_i | \Sigma_{\mathrm{xc}}(\epsilon^{\mathrm{QP}}_i) - V_{\mathrm{xc}} |
  \psi^{\mathrm{KS}}_i \rangle$. 
We note that Eq.~\ref{eq:staticsigma} has also been used in Ref.~\citenum{vanschilfgaarde_quasiparticle_2006} to implement the self-consistent
$GW$ method. 
  
\section{Results} \label{sec:results}
\subsection{Computational setup}
The electronic structure of a supercell representing a defect within a periodic
solid is initially obtained by restricted close-shell DFT calculations, with an
optimized geometry from unrestricted open-shell calculations. We use the Quantum
Espresso~\cite{Giannozzi2009} code, with the PBE ~\cite{Perdew1996} or DDH
functional~\cite{Skone2014}, SG15 norm-conserving
pseudopotentials~\cite{Schlipf2015} and a 50 Ry kinetic energy cutoff for the
plane wave basis set. Only the  $\Gamma$-point is employed to sample the
Brillouin zone of the supercell. 

The selection of the defect orbitals defining the active space may be performed
by manually identifying a set of KS eigenstates localized around the defect of
interest ~\cite{ma2020,ma2020a,ma2021} or by using Wannier functions
~\cite{muechler_quantum_2021}. However these procedures do not offer a
systematic means to verify convergence as a function of the composition and size
of the active space.

Here we introduce a localization factor, a scalar $L_V$, associated to each KS
orbital: 
\begin{equation}
    L_V(\psi_n^{\mathrm{KS}}) =  \int_{V \subseteq \Omega} {|\psi_n^{\mathrm{KS}}(\bm{x})|}^2 \mathrm{d}\bm{x}, \label{eq:LF}
\end{equation}
where $V$ is a chosen volume including the defect, smaller than the supercell
volume $\Omega$. The value of $L_V$ varies between 0 and 1. The active space for
a given defect is then defined by those  KS orbitals for which $L_V$ is larger
than a given threshold. Decreasing the value of the threshold  allows for a systematic change in  the composition and number of orbitals belonging to the
active space.

In our calculations, the parameters of the effective Hamiltonian are obtained
using constrained RPA (cRPA) calculations with the double counting correction of
Eq. \ref{eq:teff_final}, called  EDC@$G_0 W_0$.  The number of eigenpotentials
$N_{\mathrm{PDEP}}$ used for the spectral decomposition of the polarizability is
set to $512$ in all calculations. Convergence tests as a function of
$N_{\mathrm{PDEP}}$ are presented in Sec. 6 of the SI. Eigenvalues and
eigenvectors of the active-space Hamiltonian are obtained with
full-configuration interaction (FCI) calculations as implemented in the
PySCF~\cite{sun2018} code.

\begin{figure}[t!]
\includegraphics{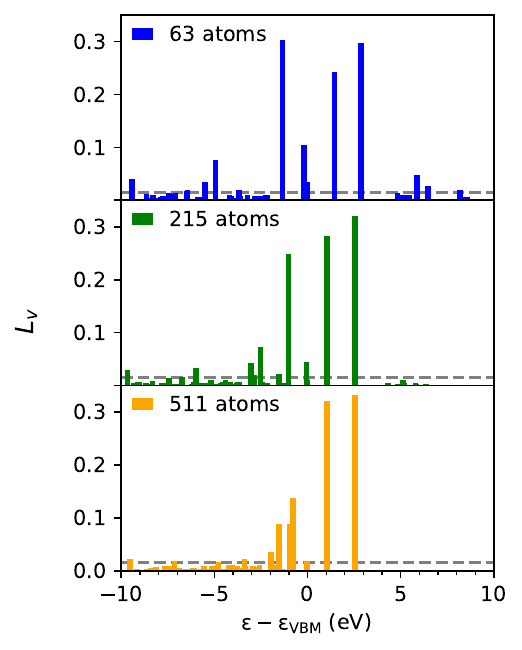}
\caption{\label{fig:nv-_localization_pbe} 
Localization factor ($L_V$, see Eq. \ref{eq:LF}) as a function of the energy of
  the Kohn-Sham orbitals, relative to the energy of the valence band maximum
  (VBM), for a \ce{NV-} center in diamond. We present results for supercells of
  three different sizes. The threshold used to define the active space is 5\%
  (see text).
}
\end{figure}
\subsection{Negatively-charged nitrogen vacancy center in diamond} 
As a prototypical spin qubit for quantum information
science\cite{Davies1976,Rogers2008,goldman2015}, the \ce{NV-} center in diamond
has been extensively studied on different levels of
theory\cite{Doherty2011,Maze2011,Choi2012,bockstedte_ab_2018,ma2020a,ma2020}. 
It is generally recognized~\cite{loubser1978,Doherty2013} that the four dangling
bonds around the defect form a minimal model for the active space, with two
non-degenerate $a_1$ orbitals, and two degenerate orbitals with $e$ character.

\begin{figure}[t!]
\includegraphics{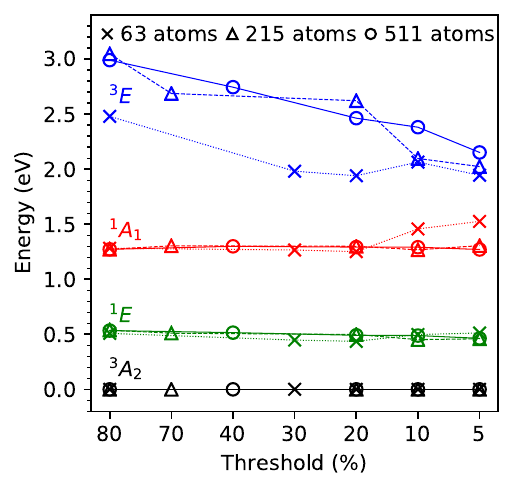}
\caption{\label{fig:nv-_active_space_pbe} 
Computed vertical excitation energies of the \ce{NV-} center in diamond as a
  function of the chosen threshold of the localization factor $L_V$ (see Eq.
  \ref{eq:LF}). States are labeled using the irreducible representation of the
  $C_{3v}$ point group. We present results obtained with the PBE functional for
  cells of three different sizes. We note that 80\% corresponds to a (4o, 6e), (3o, 4e) or (3o, 4e) active space for a 63-, 215- or 511-atom supercell respectively, and 5\% corresponds to a (22o, 42e), (14o, 26e) or (12o, 22e) active space for a 63-, 215- or 511-atom supercell respectively.
}
\end{figure}
Instead of constructing a model with a priori knowledge of the defect electronic
structure, we determine the active-space composition and size with the help of
the localization factor defined in Eq. \ref{eq:LF}, as shown in Fig.
\ref{fig:nv-_localization_pbe}. When using 63-, 215- and 511-atom supercells,
irrespective of the threshold used to define $L_V$, we find three defect
orbitals with energies within the band gap of diamond, corresponding to the two
degenerate $e$ orbitals and to one of the $a_1$ orbital of the minimal model.
In the  63- and 215-atom supercells, we find that one of the  $a_1$ orbitals
belonging to the minimal model is below the VBM of diamond;  in the 511-atom
cell we find instead that three localized orbitals are below the VBM, indicating
that at least  6 orbitals are  required to define the active space. This
suggests that the minimal model with a fixed number of orbitals may be
insufficient to accurately describe the system with a large supercell. 

\begin{figure}[t!]
\includegraphics{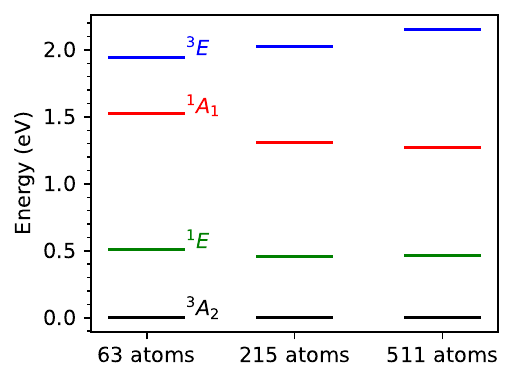}
\caption{\label{fig:nv-_supercell_pbe}
Computed vertical excitation energies for the \ce{NV-} in diamond. We show
  converged results (see Fig. \ref{fig:nv-_active_space_pbe}) at 5\%
  localization threshold as a function of the supercell size, obtained with the
  PBE functional. 
}
\end{figure}
In Fig. \ref{fig:nv-_active_space_pbe} we show the vertical excitation energies
of the \ce{NV-} center obtained by diagonalizing the effective Hamiltonian (Eq.
\ref{eq:def_hameff}), as a function of the localization threshold chosen to
define $L_V$.  Irrespective of the chosen threshold,  we  find that the ground
state has  ${}^3A_1$ symmetry. We also find that the lowest excited states with
${}^1 E$ and ${}^1 A_1$ symmetry converge much faster as a function of the $L_V$
threshold than ${}^3 E$, since ${}^3 E$ arises from  $a_1^1e^3$ configurations
rather than a $a_1^2e^2$ configurations, as in the case  of ${}^3 A_2$, ${}^1 E$
and ${}^1 A_1$. Most orbitals added to the active space when decreasing the
localization threshold  exhibit $a_1$ character.  We note that the convergence
of the ${}^1 A_1$ state in the  63-atom cell is not smooth, probably due to
orbitals with $e$ character being part of the active space as the threshold
value is decreased. Overall our results point at the need to converge the
composition and size of the active space both as a function of $L_V$ and cell
size.  In Fig. \ref{fig:nv-_supercell_pbe} we show the vertical excitation
energies of the \ce{NV-} center as a function of supercell size, and the values
are summarized in Tab. \ref{tab:nv-_excitation_energies}. We chose a 5\% $L_V$
threshold, i.e. all orbitals with $L_V \geq 0.05$ are included in the active
space $A$.  As shown in Tab.~\ref{tab:nv-_excitation_energies}, results obtained
using the  EDC@$G_0 W_0$ correction are much closer to the experimental values than those
computed using the original Hartree-Fock double counting correction (see
Section~\ref{sec:dc}), which we call here HFDC. Furthermore, we find  unphysical
excitations  (i.e. states that do not have any experimental counterpart)  with
HFDC; however such unphysical states  are not present when we use the EDC@$G_0 W_0$
correction (see  Sec. 7.3 of the SI for details).

\begin{figure}[t!]
\includegraphics{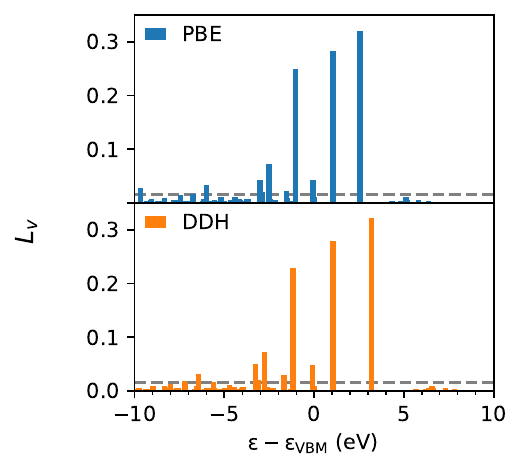}
\caption{\label{fig:nv-_localization_xc_216} Localization factor ($L_V$, see Eq.
  \ref{eq:LF}) as a function of the energy of the Kohn-Sham orbitals, relative
  to the energy of the valence band maximum (VBM), for a \ce{NV-} center in
  diamond. We present results for 215-atom cell obtained with the PBE and DDH
  functionals. The threshold used to define the active space is 5\% (see text).
}
\end{figure}

\begin{figure}[t!]
\includegraphics{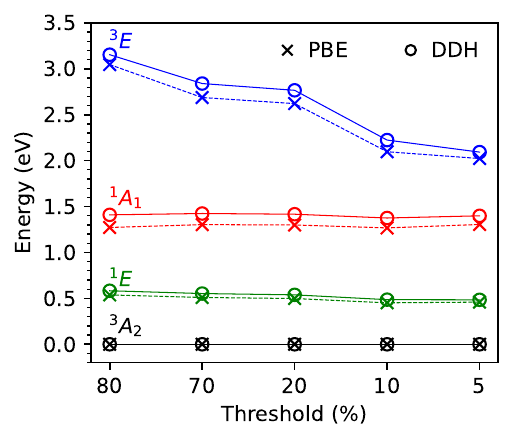}
\caption{\label{fig:nv-_active_space_xc_216} Computed vertical excitation
  energies of the \ce{NV-} center in diamond as a function of the chosen
  threshold of the localization factor $L_V$ (see Eq. \ref{eq:LF}) to define the
  active space. States are labeled using the irreducible representation of the
  $C_{3v}$ point group. We present results for 215-atom cell obtained with the
  PBE and DDH functionals. We note that 80\% threshold corresponds to a (3o, 4e) active space, and 5\% threshold corresponds to a (14o, 26e) or (15o, 28e) active space for PBE or DDH respectively.
}
\end{figure}

In our previous work, using HFDC corrrections we found substantial differences
between results obtained with the PBE or the hybrid functional
DDH\cite{Skone2014,Skone2016,Brawand2016,Brawand2017,Gerosa_2017,Zheng2019}.
Hence we analyze the influence of the chosen functional when using the EDC@$G_0 W_0$
correction. In Fig. \ref{fig:nv-_localization_xc_216} and
\ref{fig:nv-_active_space_xc_216}, we compare  PBE and DDH results for the
\ce{NV-} center, for converged active space in a 215-atom cell.  Our results
indicate that, except for a widening of the bandgap, the electronic structure is
almost insensitive to the choice of the functional. The order of localized
defect states within the gap and their localization properties are nearly
identical when using  PBE and DDH, and the shift of the position of the defect
orbitals relative to the band edges is mostly due to the difference in the PBE
and DDH bandgaps. The excitation energies of DDH calculations are less than 0.1
eV higher than their PBE counterparts. It is reasonable to expect that the
insensitivity to the functional found here in the case of the \ce{NV-} center
may apply more generally to other classes of covalently bonded semiconductors;
it appears that the sensitivity observed with the HFDC  scheme may have been
caused by the incomplete double counting correction of the DFT
exchange-correlation effects. However obtaining results for additional defects
and solids will be necessary to come to a firm conclusion.

\begin{table*}[t!]
\caption{Computed vertical excitation energies (eV) for the \ce{NV-} in diamond
  for three states (see Fig. \ref{fig:nv-_supercell_pbe}) obtained with QDET,
  the PBE functional and 511-atom supercells. We show results using the
  Hartree-Fock double counting (HFDC) and exact double counting (EDC@$G_0 W_0$) schemes
  (see text).  We also report experimental results (Exp), including results for
  zero-phonon lines (ZPL), and results obtained using $GW$ and the
  Bethe-Salpeter Equation (BSE), model fit from $GW$ solved by configuration
  interaction (CI), model obtained from constrained random phase approximation
  (cRPA) solved by CI, and quantum chemistry results on clusters from complete
  active space self-consistent field (CASSCF), multireference configuration
  interaction (MRCI) and Monte Carlo configuration interaction (MCCI).}
\begin{tabular}{llllll}
\hline
& Reference$\backslash$Electronic States & ${}^1E_{}$ & ${}^1A_{1}$ & ${}^3E_{}$  \\
\hline
& Exp\cite{Davies1976} & & & 2.18 \\
& Exp ZPL\cite{Davies1976,Rogers2008,kehayias2013,goldman2015,goldman2015a} & 0.34–0.43 & 1.51–1.60 & 1.945 \\
& QDET (EDC@$G_0 W_0$) & 0.463 & 1.270 & 2.152 \\
& QDET (HFDC) & 0.375 & 1.150 & 1.324 \\
& $GW$ + BSE\cite{ma2010} & 0.40 & 0.99 &  2.32 \\
& Model fit from $GW$ + CI\cite{Choi2012} & 0.5 & 1.5 & 2.1 \\
& Model from CRPA + CI\cite{bockstedte_ab_2018} & 0.49 & 1.41 & 2.02  \\
& \ce{C_{85} H_{76} N^{-}} CASSCF(6,6)\cite{bhandari2021} & 0.25 & 1.60 & 2.14 \\
& \ce{C_{49} H_{52} N^{-}} CASSCF(6,8)\cite{lin2008} &  &  &  2.57 \\
& \ce{C_{19} H_{28} N^{-}} MRCI(8,10)\cite{zyubin2009} & 0.50 & 1.23  & 1.36 \\
& \ce{C_{42} H_{42} N^{-}} MCCI\cite{delaney2010} & 0.63 &  2.06 &  1.96 \\
\hline
\end{tabular}
\label{tab:nv-_excitation_energies}
\end{table*}

\subsection{Neutral group-\texorpdfstring{\ce{IV}}{} vacancy centers in diamond}
\begin{figure}[t!]
\includegraphics{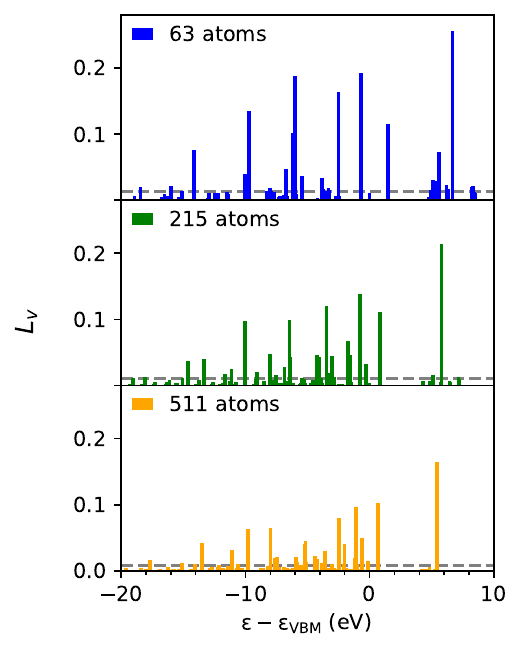}
\caption{\label{fig:siv0_localization_pbe}
Localization factor ($L_V$, see Eq. \ref{eq:LF}) as a function of the energy of
  the Kohn-Sham orbitals, relative to the energy of the valence band maximum
  (VBM), for a \ce{SiV^0} center in diamond. We present results for supercells
  of three different sizes. The threshold used to define the active space is 5\%
  (see text).
}
\end{figure}
In the last decade, a number of studies have  investigated group-\ce{IV} vacancy
centers in
diamond\cite{Gali2013,Thiering2018,Green2019,Thiering2019,zhang2020a,ma2020a,ma2020},
using either a   four-orbital\cite{Thiering2019} or a nine-orbital minimal
model\cite{ma2020}.

Similar to the case of the \ce{NV-} in diamond, we determine the active space
using the  localization factor shown in Fig. \ref{fig:siv0_localization_pbe}.
We find a considerable number of localized orbitals.  We exclude from the active
space the localized conduction band orbitals, which are around 5 eV in
\ce{SiV^0},  since we found  their contribution to the excitation energies to be
negligible (see Sec. 6 of the SI). We also exclude the defect atom's
strongly-bound atomic orbitals, present at about -20 eV in \ce{GeV^0},
\ce{SnV^0} and \ce{PbV^0}, which have almost no hybridization with the host
orbitals. 

\begin{figure}[t!]
\includegraphics{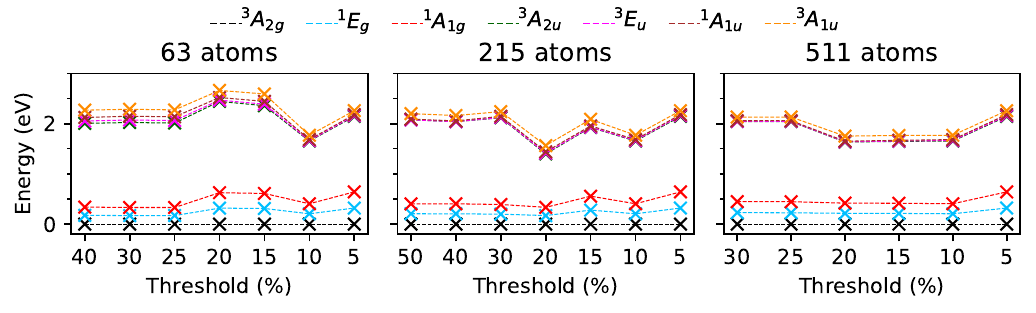}
\caption{\label{fig:siv0_active_space_pbe}
Computed vertical excitation energies of the \ce{SiV^0} center in diamond as a
  function of the chosen threshold of the localization factor $L_V$ (see Eq.
  \ref{eq:LF}) to define the active space. States are labeled using the
  irreducible representation of the $D_{3d}$ point group. We present results
  obtained with the PBE functional for cells of three different sizes. We note that a 40\%, 50\% or 30\% threshold corresponds to a (7o, 12e), (5o, 8e) or (9o, 16e) active space for a 63-, 215- or 511-atom supercell respectively, and 5\% threshold corresponds to a (32o, 62e), (40o, 78e) or (48o, 94e) active space for a 63-, 215- or 511-atom supercell respectively.
}

\end{figure}
\begin{figure}[t!]
\includegraphics{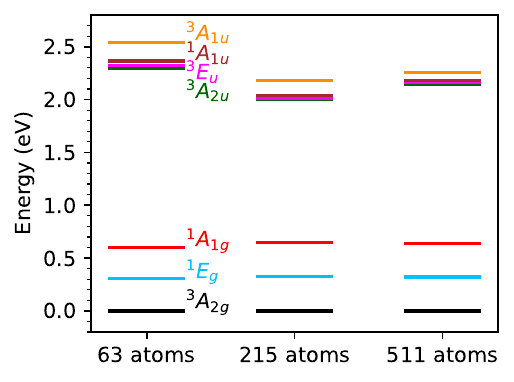}
\caption{\label{fig:siv0_supercell_pbe} Computed vertical excitation energies
  for the \ce{SiV^0} in diamond. We show converged results (see Fig.
  \ref{fig:nv-_active_space_pbe}) at 5\% localization threshold, as a function
  of the supercell size, obtained with the PBE functional.}
\end{figure}
\begin{figure}[h]
\includegraphics{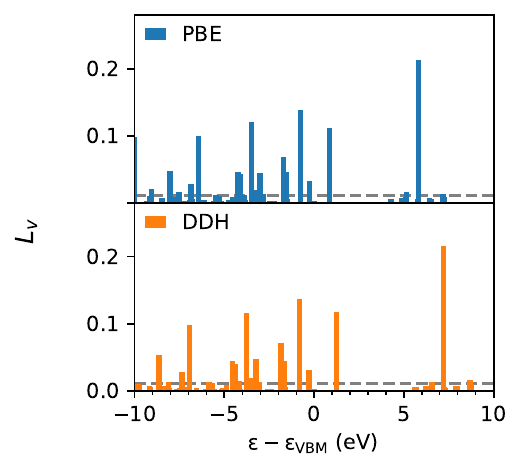}
\caption{\label{fig:siv0_localization_xc_216}
Localization factor ($L_V$, see Eq. \ref{eq:LF}) as a function of the energy of
  the Kohn-Sham orbitals, relative to the energy of the valence band maximum
  (VBM), for a \ce{SiV^0} center in diamond. We present results for a 215-atom
  cell obtained with the PBE and DDH functionals. The threshold used to define
  the active space is 5\% (see text).
}
\end{figure}
\begin{figure}[h]
\includegraphics{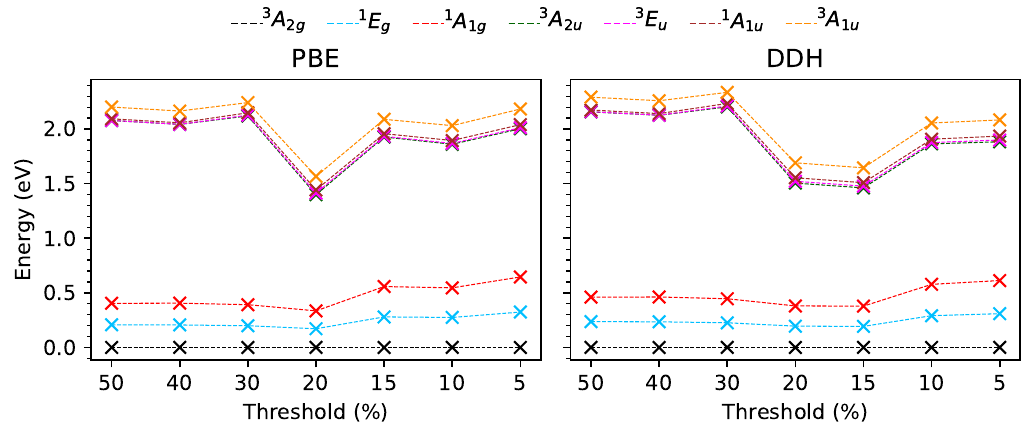}
\caption{\label{fig:siv0_active_space_xc_216}
Computed vertical excitation energies (eV) of the \ce{NV-} center in diamond as
  a function of the chosen threshold of the localization factor $L_V$ (see Eq.
  \ref{eq:LF}). States are labeled using the irreducible representation of the
  $D_{3d}$ point group. We present results for a 215-atom cell obtained with the
  PBE and DDH functionals.
}
\end{figure}

The vertical excitation energies of \ce{SiV^0} as a  function of the
localization  threshold is  reported in Fig. \ref{fig:siv0_active_space_pbe}. In
all three supercells we  find a  slow convergence of the excitation energies,
indicating that the excited states of this system are the result of the
combination of many single-particle orbitals, and that a minimal model may be
insufficient to obtain reliable excitation energies.  Using the converged
excitation energies for a given supercell, we show the convergence with
supercell size in Fig.~\ref{fig:siv0_supercell_pbe}. Similar to the case of the
\ce{NV^-} center, the low-energy excitations are well converged with a 63-atom
supercell, while the convergence of the $^3A_{2u}$, $^3E_u$, $^1A_{1u}$, and
$^3A_{1u}$ states is slower. In Tab. \ref{tab:siv0_excitation_energies}, we
compare our best converged values obtained with the EDC@$G_0 W_0$ correction with those
obtained with the HFDC correction, as well as with available experimental and
theoretical data. In general, the energies predicted using EDC@$G_0 W_0$ are higher than
those obtained with  HFDC, and in better agreement with those of  quantum
chemical cluster calculations\cite{mitra2021}. We note that the experimental
zero phonon line (ZPL) corresponding to the ${}^3E_u$ level is 1.31 eV, but the
contribution from the dynamical Jahn-Teller effect is  unknown.  Furthermore,
the excitation energies computed with  EDC@$G_0 W_0$ show faster convergence compared to
those with HFDC. For example, using EDC@$G_0 W_0$ (HFDC)  we find a difference of   0.15
(0.65) eV with  63-atom and 215-atom cells.  As shown in Figs.
\ref{fig:siv0_localization_xc_216} and \ref{fig:siv0_active_space_xc_216}, our
results with the EDC@$G_0 W_0$  scheme showed insensitivity to the choice of the
functional. Our results for \ce{GeV^0}, \ce{SnV^0} and \ce{PbV^0} are similar to
those of \ce{SiV^0} and  are summarized in Tab.
\ref{tab:siv0_excitation_energies} as well as Tab.
S1 in the SI.

\begin{table*}[t!]
\caption{Computed vertical excitation energies (eV) for the \ce{SiV^0},
  \ce{GeV^0}, \ce{SnV^0} and \ce{PbV^0} in diamond for six states (see also Fig.
  \ref{fig:siv0_supercell_pbe} for the \ce{SiV^0} results) obtained with QDET,
  the PBE functional and 511-atom supercells. We show results using the
  Hartree-Fock double counting (HFDC) and exact double counting (EDC@$G_0 W_0$) schemes
  (see text).  We also report experimental results for zero-phonon lines (ZPL),
  and results obtained with a combination of second-order $N$-electron valence
  state perturbation theory (NEVPT2) and density matrix embedding theory (DMET),
  and quantum chemistry calculations on clusters from NEVPT2. }
\begin{tabular}{lllllllll}
\hline
& System & Reference$\backslash$Electronic States & ${}^1E_{g}$ & ${}^1A_{1g}$ & ${}^3A_{2u}$ & ${}^3E_{u}$ & ${}^1A_{1u}$ & ${}^3A_{1u}$ \\
\hline
& \multirow{5}{*}{\ce{SiV^0}} & Exp ZPL & & & & 1.31 & & \\
& & QDET (EDC@$G_0 W_0$) & 0.321 & 0.642 & 2.146 & 2.161 & 2.183 & 2.260 \\
& & QDET (HFDC) & 0.236 & 0.435 & 1.098 & 1.096 & 1.111 & 1.188 \\
& & NEVPT2-DMET(10,12)\cite{mitra2021} & 0.51 & 1.14 & 2.39 & 2.47 & & 2.61 \\
& & \ce{C_{84} H_{78} Si_0 } NEVPT2(10,12)\cite{mitra2021} & 0.54 & 1.10 & 2.10 & 2.16 & & 2.14 \\
\hline
& \multirow{2}{*}{\ce{GeV^0}} & QDET (EDC@$G_0 W_0$) & 0.357 & 0.717 & 2.924 & 2.925 & 2.940 & 2.970 \\
& & QDET (HFDC) & 0.289 & 0.554 & 1.456 & 1.443 & 1.443 & 1.495 \\
\hline
& \multirow{2}{*}{\ce{SnV^0}} & QDET (EDC@$G_0 W_0$) & 0.295 & 0.596 & 2.590 & 2.571 & 2.561 & 2.616 \\
& & QDET (HFDC) & 0.276 & 0.551 & 1.459 & 1.444 & 1.436 & 1.491  \\
\hline
& \multirow{2}{*}{\ce{PbV^0}} & QDET (EDC@$G_0 W_0$) & 0.319 & 0.640 & 3.095 & 3.072 & 3.056 & 3.099 \\
& & QDET (HFDC) & 0.302 & 0.600 & 1.788 & 1.768 & 1.755 & 1.796 \\
\hline
\end{tabular}
\label{tab:siv0_excitation_energies}
\end{table*}

\section{Conclusions} \label{sec:conclusions}
In summary, in this work we presented a Green’s function formulation of the
quantum defect embedding theory (QDET) that enables the definition of an
improved correction to the double counting scheme originally adopted in
Refs~\citenum{ma2020a,ma2021}. 
We defined an effective Hamiltonian for the active space  within a Green's function formalism, where the effective interaction is static and the self-energy cross-terms between the active space and the environment are neglected. Our results show that these approximations are  appropriate  to describe the localized defect states in semiconductors investigated in this work. Within the Green's function formalism adopted here, we derived an exact double counting scheme (EDC@$G_0W_0$) replacing the approximate scheme originally adopted in
Refs~\citenum{ma2020a,ma2021}. We emphasize that the double counting correction EDC@$G_0W_0$ enables the removal of any double counting terms arising from the separation of the whole system into active space and environment.
We then described the implementation of the scheme within the
WEST code~\cite{govoni2015}, including a strategy to ensure convergence of our
calculations with respect to the size and composition of the active space.
Further, we demonstrated that QDET with exact double counting provides reliable
results for several defects in diamond, with negligible dependence on the
functional chosen for the  underlying DFT calculations of the defects. Work is
in progress to apply QDET with the EDC@$G_0 W_0$ scheme to more complex systems, such as
defects in oxides and molecules on surfaces.

\section*{Supporting information}
Definition of matrix elements (section 1), comparison with other double counting
schemes in the literature (section 2), implementation details (section 3),
Hartree-Fock double counting (section 4), convergence of QDET calculations
(section 5), discussion of ghost states (section 6), and a table of vertical
excitation energies (section 7).

\section*{Acknowledgements}

This work was supported by MICCoM, as part of the Computational Materials
Sciences Program funded by the U.S. Department of Energy, Office of Science,
Basic Energy Sciences, Materials Sciences and Engineering Division through
Argonne National Laboratory, under contract number DE-AC02-06CH11357. This
research used resources of the National Energy Research Scientific Computing
Center (NERSC), a DOE Office of Science User Facility supported by the Office of
Science of the US Department of Energy under Contract No. DE-AC02-05CH11231, and
resources of the University of Chicago Research Computing Center.

\bibliography{ref}

\providecommand{\latin}[1]{#1}
\makeatletter
\providecommand{\doi}
  {\begingroup\let\do\@makeother\dospecials
  \catcode`\{=1 \catcode`\}=2 \doi@aux}
\providecommand{\doi@aux}[1]{\endgroup\texttt{#1}}
\makeatother
\providecommand*\mcitethebibliography{\thebibliography}
\csname @ifundefined\endcsname{endmcitethebibliography}
  {\let\endmcitethebibliography\endthebibliography}{}
\begin{mcitethebibliography}{81}
\providecommand*\natexlab[1]{#1}
\providecommand*\mciteSetBstSublistMode[1]{}
\providecommand*\mciteSetBstMaxWidthForm[2]{}
\providecommand*\mciteBstWouldAddEndPuncttrue
  {\def\EndOfBibitem{\unskip.}}
\providecommand*\mciteBstWouldAddEndPunctfalse
  {\let\EndOfBibitem\relax}
\providecommand*\mciteSetBstMidEndSepPunct[3]{}
\providecommand*\mciteSetBstSublistLabelBeginEnd[3]{}
\providecommand*\EndOfBibitem{}
\mciteSetBstSublistMode{f}
\mciteSetBstMaxWidthForm{subitem}{(\alph{mcitesubitemcount})}
\mciteSetBstSublistLabelBeginEnd
  {\mcitemaxwidthsubitemform\space}
  {\relax}
  {\relax}

\bibitem[Hohenberg and Kohn(1964)Hohenberg, and Kohn]{hohenberg1964}
Hohenberg,~P.; Kohn,~W. Inhomogeneous Electron Gas. \emph{Phys.~Rev.~}
  \textbf{1964}, \emph{136}, B864--B871\relax
\mciteBstWouldAddEndPuncttrue
\mciteSetBstMidEndSepPunct{\mcitedefaultmidpunct}
{\mcitedefaultendpunct}{\mcitedefaultseppunct}\relax
\EndOfBibitem
\bibitem[Kohn and Sham(1965)Kohn, and Sham]{kohn1965}
Kohn,~W.; Sham,~L.~J. Self-Consistent Equations Including Exchange and
  Correlation Effects. \emph{Phys.~Rev.~} \textbf{1965}, \emph{140},
  A1133--A1138\relax
\mciteBstWouldAddEndPuncttrue
\mciteSetBstMidEndSepPunct{\mcitedefaultmidpunct}
{\mcitedefaultendpunct}{\mcitedefaultseppunct}\relax
\EndOfBibitem
\bibitem[Neugebauer and Hickel(2013)Neugebauer, and Hickel]{neugebauer2013}
Neugebauer,~J.; Hickel,~T. Density Functional Theory in Materials Science.
  \emph{WIREs Computational Molecular Science} \textbf{2013}, \emph{3},
  438--448\relax
\mciteBstWouldAddEndPuncttrue
\mciteSetBstMidEndSepPunct{\mcitedefaultmidpunct}
{\mcitedefaultendpunct}{\mcitedefaultseppunct}\relax
\EndOfBibitem
\bibitem[Mardirossian and {Head-Gordon}(2017)Mardirossian, and
  {Head-Gordon}]{mardirossian2017}
Mardirossian,~N.; {Head-Gordon},~M. Thirty Years of Density Functional Theory
  in Computational Chemistry: An Overview and Extensive Assessment of 200
  Density Functionals. \emph{Mol.~Phys.} \textbf{2017}, \emph{115},
  2315--2372\relax
\mciteBstWouldAddEndPuncttrue
\mciteSetBstMidEndSepPunct{\mcitedefaultmidpunct}
{\mcitedefaultendpunct}{\mcitedefaultseppunct}\relax
\EndOfBibitem
\bibitem[Jones(2015)]{jones2015}
Jones,~R.~O. Density Functional Theory: {{Its}} Origins, Rise to Prominence,
  and Future. \emph{Rev.~Mod.~Phys.~} \textbf{2015}, \emph{87}, 897--923\relax
\mciteBstWouldAddEndPuncttrue
\mciteSetBstMidEndSepPunct{\mcitedefaultmidpunct}
{\mcitedefaultendpunct}{\mcitedefaultseppunct}\relax
\EndOfBibitem
\bibitem[Burke(2012)]{burke2012}
Burke,~K. Perspective on density functional theory. \emph{J.~Chem.~Phys.~}
  \textbf{2012}, \emph{136}, 150901\relax
\mciteBstWouldAddEndPuncttrue
\mciteSetBstMidEndSepPunct{\mcitedefaultmidpunct}
{\mcitedefaultendpunct}{\mcitedefaultseppunct}\relax
\EndOfBibitem
\bibitem[Strinati(1988)]{strinati1988}
Strinati,~G. Application of the green’s functions method to the study of the
  optical properties of semiconductors. \emph{Riv.~Nuovo~Cimento~}
  \textbf{1988}, \emph{11}, 1--86\relax
\mciteBstWouldAddEndPuncttrue
\mciteSetBstMidEndSepPunct{\mcitedefaultmidpunct}
{\mcitedefaultendpunct}{\mcitedefaultseppunct}\relax
\EndOfBibitem
\bibitem[Martin \latin{et~al.}(2016)Martin, Reining, and Ceperley]{martin2016}
Martin,~R.~M.; Reining,~L.; Ceperley,~D.~M. \emph{Interacting electrons};
  Cambridge University Press, 2016\relax
\mciteBstWouldAddEndPuncttrue
\mciteSetBstMidEndSepPunct{\mcitedefaultmidpunct}
{\mcitedefaultendpunct}{\mcitedefaultseppunct}\relax
\EndOfBibitem
\bibitem[Golze \latin{et~al.}(2019)Golze, Dvorak, and Rinke]{golze2019}
Golze,~D.; Dvorak,~M.; Rinke,~P. The {GW} Compendium: A Practical Guide to
  Theoretical Photoemission Spectroscopy. \emph{Front.~Chem.} \textbf{2019},
  \emph{7}, 377\relax
\mciteBstWouldAddEndPuncttrue
\mciteSetBstMidEndSepPunct{\mcitedefaultmidpunct}
{\mcitedefaultendpunct}{\mcitedefaultseppunct}\relax
\EndOfBibitem
\bibitem[Bartlett and Musia{\l}(2007)Bartlett, and Musia{\l}]{bartlett2007}
Bartlett,~R.~J.; Musia{\l},~M. Coupled-Cluster Theory in Quantum Chemistry.
  \emph{Rev.~Mod.~Phys.~} \textbf{2007}, \emph{79}, 291--352\relax
\mciteBstWouldAddEndPuncttrue
\mciteSetBstMidEndSepPunct{\mcitedefaultmidpunct}
{\mcitedefaultendpunct}{\mcitedefaultseppunct}\relax
\EndOfBibitem
\bibitem[Zhang and Gr{\"u}neis(2019)Zhang, and Gr{\"u}neis]{zhang2019}
Zhang,~I.~Y.; Gr{\"u}neis,~A. Coupled {{Cluster Theory}} in {{Materials
  Science}}. \emph{Front.~Mater.} \textbf{2019}, \emph{6}\relax
\mciteBstWouldAddEndPuncttrue
\mciteSetBstMidEndSepPunct{\mcitedefaultmidpunct}
{\mcitedefaultendpunct}{\mcitedefaultseppunct}\relax
\EndOfBibitem
\bibitem[Helgaker \latin{et~al.}(2014)Helgaker, Jorgensen, and
  Olsen]{helgaker2014}
Helgaker,~T.; Jorgensen,~P.; Olsen,~J. \emph{Molecular electronic-structure
  theory}; John Wiley \& Sons, 2014\relax
\mciteBstWouldAddEndPuncttrue
\mciteSetBstMidEndSepPunct{\mcitedefaultmidpunct}
{\mcitedefaultendpunct}{\mcitedefaultseppunct}\relax
\EndOfBibitem
\bibitem[Sun and Chan(2016)Sun, and Chan]{sun2016}
Sun,~Q.; Chan,~G. K.-L. Quantum Embedding Theories. \emph{Acc.~Chem.~Res.~}
  \textbf{2016}, \emph{49}, 2705--2712\relax
\mciteBstWouldAddEndPuncttrue
\mciteSetBstMidEndSepPunct{\mcitedefaultmidpunct}
{\mcitedefaultendpunct}{\mcitedefaultseppunct}\relax
\EndOfBibitem
\bibitem[Jones \latin{et~al.}(2020)Jones, Mosquera, Schatz, and
  Ratner]{jones2020}
Jones,~L.~O.; Mosquera,~M.~A.; Schatz,~G.~C.; Ratner,~M.~A. Embedding Methods
  for Quantum Chemistry: Applications from Materials to Life Sciences.
  \emph{J.~Am.~Chem.~Soc.~} \textbf{2020}, \emph{142}, 3281--3295\relax
\mciteBstWouldAddEndPuncttrue
\mciteSetBstMidEndSepPunct{\mcitedefaultmidpunct}
{\mcitedefaultendpunct}{\mcitedefaultseppunct}\relax
\EndOfBibitem
\bibitem[Sheng \latin{et~al.}(2021)Sheng, Vorwerk, Govoni, and
  Galli]{sheng_quantum_2021}
Sheng,~N.; Vorwerk,~C.; Govoni,~M.; Galli,~G. Quantum {Simulations} of
  {Material} {Properties} on {Quantum} {Computers}. \emph{arXiv:2105.04736
  [cond-mat, physics:quant-ph]} \textbf{2021}, arXiv: 2105.04736\relax
\mciteBstWouldAddEndPuncttrue
\mciteSetBstMidEndSepPunct{\mcitedefaultmidpunct}
{\mcitedefaultendpunct}{\mcitedefaultseppunct}\relax
\EndOfBibitem
\bibitem[Senn and Thiel(2009)Senn, and Thiel]{senn2009}
Senn,~H.~M.; Thiel,~W. {{QM}}/{{MM Methods}} for {{Biomolecular Systems}}.
  \emph{Angew.~Chem.~Int.~Ed.~} \textbf{2009}, \emph{48}, 1198--1229\relax
\mciteBstWouldAddEndPuncttrue
\mciteSetBstMidEndSepPunct{\mcitedefaultmidpunct}
{\mcitedefaultendpunct}{\mcitedefaultseppunct}\relax
\EndOfBibitem
\bibitem[Liu \latin{et~al.}(2014)Liu, Wang, Chen, Field, and Gao]{liu2014}
Liu,~M.; Wang,~Y.; Chen,~Y.; Field,~M.~J.; Gao,~J. {{QM}}/{{MM}} through the
  1990s: {{The First Twenty Years}} of {{Method Development}} and
  {{Applications}}. \emph{Isr.~J.~Chem.} \textbf{2014}, \emph{54},
  1250--1263\relax
\mciteBstWouldAddEndPuncttrue
\mciteSetBstMidEndSepPunct{\mcitedefaultmidpunct}
{\mcitedefaultendpunct}{\mcitedefaultseppunct}\relax
\EndOfBibitem
\bibitem[Libisch \latin{et~al.}(2014)Libisch, Huang, and Carter]{libisch2014}
Libisch,~F.; Huang,~C.; Carter,~E.~A. Embedded Correlated Wavefunction Schemes:
  Theory and Applications. \emph{Acc.~Chem.~Res.~} \textbf{2014}, \emph{47},
  2768--2775\relax
\mciteBstWouldAddEndPuncttrue
\mciteSetBstMidEndSepPunct{\mcitedefaultmidpunct}
{\mcitedefaultendpunct}{\mcitedefaultseppunct}\relax
\EndOfBibitem
\bibitem[Gomes \latin{et~al.}(2008)Gomes, Jacob, and Visscher]{gomes2008}
Gomes,~A. S.~P.; Jacob,~C.~R.; Visscher,~L. Calculation of Local Excitations in
  Large Systems by Embedding Wave-Function Theory in Density-Functional Theory.
  \emph{Phys.~Chem.~Chem.~Phys.~} \textbf{2008}, \emph{10}, 5353--5362\relax
\mciteBstWouldAddEndPuncttrue
\mciteSetBstMidEndSepPunct{\mcitedefaultmidpunct}
{\mcitedefaultendpunct}{\mcitedefaultseppunct}\relax
\EndOfBibitem
\bibitem[Goodpaster \latin{et~al.}(2012)Goodpaster, Barnes, Manby, and
  Miller]{goodpaster2012}
Goodpaster,~J.~D.; Barnes,~T.~A.; Manby,~F.~R.; Miller,~T.~F. Density
  Functional Theory Embedding for Correlated Wavefunctions: {{Improved}}
  Methods for Open-Shell Systems and Transition Metal Complexes.
  \emph{J.~Chem.~Phys.~} \textbf{2012}, \emph{137}, 224113\relax
\mciteBstWouldAddEndPuncttrue
\mciteSetBstMidEndSepPunct{\mcitedefaultmidpunct}
{\mcitedefaultendpunct}{\mcitedefaultseppunct}\relax
\EndOfBibitem
\bibitem[Wouters \latin{et~al.}(2016)Wouters, {Jim{\'e}nez-Hoyos}, Sun, and
  Chan]{wouters2016}
Wouters,~S.; {Jim{\'e}nez-Hoyos},~C.~A.; Sun,~Q.; Chan,~G. K.-L. A Practical
  Guide to Density Matrix Embedding Theory in Quantum Chemistry.
  \emph{J.~Chem.~Theory.~Comput.~} \textbf{2016}, \emph{12}, 2706--2719\relax
\mciteBstWouldAddEndPuncttrue
\mciteSetBstMidEndSepPunct{\mcitedefaultmidpunct}
{\mcitedefaultendpunct}{\mcitedefaultseppunct}\relax
\EndOfBibitem
\bibitem[Knizia and Chan(2012)Knizia, and Chan]{knizia2012}
Knizia,~G.; Chan,~G. K.-L. Density Matrix Embedding: A Simple Alternative to
  Dynamical Mean-Field Theory. \emph{Phys.~Rev.~Lett.~} \textbf{2012},
  \emph{109}, 186404\relax
\mciteBstWouldAddEndPuncttrue
\mciteSetBstMidEndSepPunct{\mcitedefaultmidpunct}
{\mcitedefaultendpunct}{\mcitedefaultseppunct}\relax
\EndOfBibitem
\bibitem[Knizia and Chan(2013)Knizia, and Chan]{knizia2013}
Knizia,~G.; Chan,~G. K.-L. Density Matrix Embedding: A Strong-Coupling Quantum
  Embedding Theory. \emph{J.~Chem.~Theory.~Comput.~} \textbf{2013}, \emph{9},
  1428--1432\relax
\mciteBstWouldAddEndPuncttrue
\mciteSetBstMidEndSepPunct{\mcitedefaultmidpunct}
{\mcitedefaultendpunct}{\mcitedefaultseppunct}\relax
\EndOfBibitem
\bibitem[Cui \latin{et~al.}(2020)Cui, Zhu, and Chan]{cui_efficient_2020}
Cui,~Z.-H.; Zhu,~T.; Chan,~G. K.-L. Efficient Implementation of Ab Initio
  Quantum Embedding in Periodic Systems: Density Matrix Embedding Theory.
  \textbf{2020}, \emph{16}, 119--129\relax
\mciteBstWouldAddEndPuncttrue
\mciteSetBstMidEndSepPunct{\mcitedefaultmidpunct}
{\mcitedefaultendpunct}{\mcitedefaultseppunct}\relax
\EndOfBibitem
\bibitem[Pham \latin{et~al.}(2020)Pham, Hermes, and Gagliardi]{pham2020}
Pham,~H.~Q.; Hermes,~M.~R.; Gagliardi,~L. Periodic Electronic Structure
  Calculations with the Density Matrix Embedding Theory.
  \emph{J.~Chem.~Theory.~Comput.~} \textbf{2020}, \emph{16}, 130--140\relax
\mciteBstWouldAddEndPuncttrue
\mciteSetBstMidEndSepPunct{\mcitedefaultmidpunct}
{\mcitedefaultendpunct}{\mcitedefaultseppunct}\relax
\EndOfBibitem
\bibitem[Hermes and Gagliardi(2019)Hermes, and Gagliardi]{hermes2019}
Hermes,~M.~R.; Gagliardi,~L. Multiconfigurational Self-Consistent Field Theory
  with Density Matrix Embedding: The Localized Active Space Self-Consistent
  Field Method. \emph{J.~Chem.~Theory.~Comput.~} \textbf{2019}, \emph{15},
  972--986\relax
\mciteBstWouldAddEndPuncttrue
\mciteSetBstMidEndSepPunct{\mcitedefaultmidpunct}
{\mcitedefaultendpunct}{\mcitedefaultseppunct}\relax
\EndOfBibitem
\bibitem[Pham \latin{et~al.}(2018)Pham, Bernales, and Gagliardi]{pham2018}
Pham,~H.~Q.; Bernales,~V.; Gagliardi,~L. Can Density Matrix Embedding Theory
  with the Complete Activate Space Self-Consistent Field Solver Describe Single
  and Double Bond Breaking in Molecular Systems?
  \emph{J.~Chem.~Theory.~Comput.~} \textbf{2018}, \emph{14}, 1960--1968\relax
\mciteBstWouldAddEndPuncttrue
\mciteSetBstMidEndSepPunct{\mcitedefaultmidpunct}
{\mcitedefaultendpunct}{\mcitedefaultseppunct}\relax
\EndOfBibitem
\bibitem[Lan and Zgid(2017)Lan, and Zgid]{lan2017}
Lan,~T.~N.; Zgid,~D. Generalized Self-Energy Embedding Theory.
  \emph{J.~Chem.~Phys.~Lett.} \textbf{2017}, \emph{8}, 2200--2205\relax
\mciteBstWouldAddEndPuncttrue
\mciteSetBstMidEndSepPunct{\mcitedefaultmidpunct}
{\mcitedefaultendpunct}{\mcitedefaultseppunct}\relax
\EndOfBibitem
\bibitem[Zgid and Gull(2017)Zgid, and Gull]{zgid2017}
Zgid,~D.; Gull,~E. Finite Temperature Quantum Embedding Theories for Correlated
  Systems. \emph{New.~J.~Phys.~} \textbf{2017}, \emph{19}, 023047\relax
\mciteBstWouldAddEndPuncttrue
\mciteSetBstMidEndSepPunct{\mcitedefaultmidpunct}
{\mcitedefaultendpunct}{\mcitedefaultseppunct}\relax
\EndOfBibitem
\bibitem[Rusakov \latin{et~al.}(2019)Rusakov, Iskakov, Tran, and
  Zgid]{rusakov2019}
Rusakov,~A.~A.; Iskakov,~S.; Tran,~L.~N.; Zgid,~D. Self-Energy Embedding Theory
  ({{SEET}}) for Periodic Systems. \emph{J.~Chem.~Theory.~Comput.~}
  \textbf{2019}, \emph{15}, 229--240\relax
\mciteBstWouldAddEndPuncttrue
\mciteSetBstMidEndSepPunct{\mcitedefaultmidpunct}
{\mcitedefaultendpunct}{\mcitedefaultseppunct}\relax
\EndOfBibitem
\bibitem[Georges and Kotliar(1992)Georges, and Kotliar]{georges1992}
Georges,~A.; Kotliar,~G. Hubbard Model in Infinite Dimensions.
  \emph{Phys.~Rev.~B} \textbf{1992}, \emph{45}, 6479--6483\relax
\mciteBstWouldAddEndPuncttrue
\mciteSetBstMidEndSepPunct{\mcitedefaultmidpunct}
{\mcitedefaultendpunct}{\mcitedefaultseppunct}\relax
\EndOfBibitem
\bibitem[Georges \latin{et~al.}(1996)Georges, Kotliar, Krauth, and
  Rozenberg]{georges1996}
Georges,~A.; Kotliar,~G.; Krauth,~W.; Rozenberg,~M.~J. Dynamical Mean-Field
  Theory of Strongly Correlated Fermion Systems and the Limit of Infinite
  Dimensions. \emph{Rev.~Mod.~Phys.~} \textbf{1996}, \emph{68}, 13--125\relax
\mciteBstWouldAddEndPuncttrue
\mciteSetBstMidEndSepPunct{\mcitedefaultmidpunct}
{\mcitedefaultendpunct}{\mcitedefaultseppunct}\relax
\EndOfBibitem
\bibitem[Georges(2004)]{georges2004}
Georges,~A. Strongly Correlated Electron Materials: Dynamical Mean-Field Theory
  and Electronic Structure. \emph{AIP~Conf.~Proc.~} \textbf{2004}, \emph{715},
  3--74\relax
\mciteBstWouldAddEndPuncttrue
\mciteSetBstMidEndSepPunct{\mcitedefaultmidpunct}
{\mcitedefaultendpunct}{\mcitedefaultseppunct}\relax
\EndOfBibitem
\bibitem[Anisimov \latin{et~al.}(1997)Anisimov, Poteryaev, Korotin, Anokhin,
  and Kotliar]{anisimov1997}
Anisimov,~V.~I.; Poteryaev,~A.~I.; Korotin,~M.~A.; Anokhin,~A.~O.; Kotliar,~G.
  First-Principles Calculations of the Electronic Structure and Spectra of
  Strongly Correlated Systems: Dynamical Mean-Field Theory.
  \emph{J.~Phys.~Condens.~Matter.~} \textbf{1997}, \emph{9}, 7359--7367\relax
\mciteBstWouldAddEndPuncttrue
\mciteSetBstMidEndSepPunct{\mcitedefaultmidpunct}
{\mcitedefaultendpunct}{\mcitedefaultseppunct}\relax
\EndOfBibitem
\bibitem[Kotliar \latin{et~al.}(2006)Kotliar, Savrasov, Haule, Oudovenko,
  Parcollet, and Marianetti]{kotliar2006}
Kotliar,~G.; Savrasov,~S.~Y.; Haule,~K.; Oudovenko,~V.~S.; Parcollet,~O.;
  Marianetti,~C.~A. Electronic Structure Calculations with Dynamical Mean-Field
  Theory. \emph{Rev.~Mod.~Phys.~} \textbf{2006}, \emph{78}, 865--951\relax
\mciteBstWouldAddEndPuncttrue
\mciteSetBstMidEndSepPunct{\mcitedefaultmidpunct}
{\mcitedefaultendpunct}{\mcitedefaultseppunct}\relax
\EndOfBibitem
\bibitem[Ma \latin{et~al.}(2020)Ma, Govoni, and Galli]{ma2020a}
Ma,~H.; Govoni,~M.; Galli,~G. Quantum Simulations of Materials on Near-Term
  Quantum Computers. \emph{npj~Comput.~Mater.} \textbf{2020}, \emph{6},
  1--8\relax
\mciteBstWouldAddEndPuncttrue
\mciteSetBstMidEndSepPunct{\mcitedefaultmidpunct}
{\mcitedefaultendpunct}{\mcitedefaultseppunct}\relax
\EndOfBibitem
\bibitem[Ma \latin{et~al.}(2021)Ma, Sheng, Govoni, and Galli]{ma2021}
Ma,~H.; Sheng,~N.; Govoni,~M.; Galli,~G. Quantum Embedding Theory for Strongly
  Correlated States in Materials. \emph{J.~Chem.~Theory.~Comput.~}
  \textbf{2021}, \emph{17}, 2116--2125\relax
\mciteBstWouldAddEndPuncttrue
\mciteSetBstMidEndSepPunct{\mcitedefaultmidpunct}
{\mcitedefaultendpunct}{\mcitedefaultseppunct}\relax
\EndOfBibitem
\bibitem[Mitra \latin{et~al.}(2021)Mitra, Pham, Pandharkar, Hermes, and
  Gagliardi]{mitra2021}
Mitra,~A.; Pham,~H.~Q.; Pandharkar,~R.; Hermes,~M.~R.; Gagliardi,~L. Excited
  {{States}} of {{Crystalline Point Defects}} with {{Multireference Density
  Matrix Embedding Theory}}. \emph{J.~Phys.~Chem.~Lett.} \textbf{2021},
  \emph{12}, 11688--11694\relax
\mciteBstWouldAddEndPuncttrue
\mciteSetBstMidEndSepPunct{\mcitedefaultmidpunct}
{\mcitedefaultendpunct}{\mcitedefaultseppunct}\relax
\EndOfBibitem
\bibitem[Bockstedte \latin{et~al.}(2018)Bockstedte, Schütz, Garratt, Ivády,
  and Gali]{bockstedte_ab_2018}
Bockstedte,~M.; Schütz,~F.; Garratt,~T.; Ivády,~V.; Gali,~A. Ab initio
  description of highly correlated states in defects for realizing quantum
  bits. \emph{npj~Quantum~Mater.} \textbf{2018}, \emph{3}, 1--6\relax
\mciteBstWouldAddEndPuncttrue
\mciteSetBstMidEndSepPunct{\mcitedefaultmidpunct}
{\mcitedefaultendpunct}{\mcitedefaultseppunct}\relax
\EndOfBibitem
\bibitem[Ma \latin{et~al.}(2020)Ma, Sheng, Govoni, and Galli]{ma2020}
Ma,~H.; Sheng,~N.; Govoni,~M.; Galli,~G. First-Principles Studies of Strongly
  Correlated States in Defect Spin Qubits in Diamond.
  \emph{Phys.~Chem.~Chem.~Phys.~} \textbf{2020}, \emph{22}, 25522--25527\relax
\mciteBstWouldAddEndPuncttrue
\mciteSetBstMidEndSepPunct{\mcitedefaultmidpunct}
{\mcitedefaultendpunct}{\mcitedefaultseppunct}\relax
\EndOfBibitem
\bibitem[Muechler \latin{et~al.}(2021)Muechler, Badrtdinov, Hampel, Cano,
  Rösner, and Dreyer]{muechler_quantum_2021}
Muechler,~L.; Badrtdinov,~D.~I.; Hampel,~A.; Cano,~J.; Rösner,~M.;
  Dreyer,~C.~E. Quantum embedding methods for correlated excited states of
  point defects: {Case} studies and challenges. \emph{arXiv:2105.08705
  [cond-mat]} \textbf{2021}, arXiv: 2105.08705\relax
\mciteBstWouldAddEndPuncttrue
\mciteSetBstMidEndSepPunct{\mcitedefaultmidpunct}
{\mcitedefaultendpunct}{\mcitedefaultseppunct}\relax
\EndOfBibitem
\bibitem[Pfäffle \latin{et~al.}(2021)Pfäffle, Antonov, Wrachtrup, and
  Bester]{pfaffle_screened_2021}
Pfäffle,~W.; Antonov,~D.; Wrachtrup,~J.; Bester,~G. Screened configuration
  interaction method for open-shell excited states applied to {NV} centers.
  \emph{Phys.~Rev.~B} \textbf{2021}, \emph{104}, 104105\relax
\mciteBstWouldAddEndPuncttrue
\mciteSetBstMidEndSepPunct{\mcitedefaultmidpunct}
{\mcitedefaultendpunct}{\mcitedefaultseppunct}\relax
\EndOfBibitem
\bibitem[Govoni and Galli(2015)Govoni, and Galli]{govoni2015}
Govoni,~M.; Galli,~G. Large Scale {{GW}} Calculations.
  \emph{J.~Chem.~Theory.~Comput.~} \textbf{2015}, \emph{11}, 2680--2696\relax
\mciteBstWouldAddEndPuncttrue
\mciteSetBstMidEndSepPunct{\mcitedefaultmidpunct}
{\mcitedefaultendpunct}{\mcitedefaultseppunct}\relax
\EndOfBibitem
\bibitem[Wilson \latin{et~al.}(2008)Wilson, Gygi, and
  Galli]{wilson_efficient_2008}
Wilson,~H.~F.; Gygi,~F.; Galli,~G. Efficient iterative method for calculations
  of dielectric matrices. \emph{Phys.~Rev.~B} \textbf{2008}, \emph{78},
  113303\relax
\mciteBstWouldAddEndPuncttrue
\mciteSetBstMidEndSepPunct{\mcitedefaultmidpunct}
{\mcitedefaultendpunct}{\mcitedefaultseppunct}\relax
\EndOfBibitem
\bibitem[Baroni \latin{et~al.}(2001)Baroni, {de Gironcoli}, Dal~Corso, and
  Giannozzi]{baroni2001}
Baroni,~S.; {de Gironcoli},~S.; Dal~Corso,~A.; Giannozzi,~P. Phonons and
  Related Crystal Properties from Density-Functional Perturbation Theory.
  \emph{Rev.~Mod.~Phys.~} \textbf{2001}, \emph{73}, 515--562\relax
\mciteBstWouldAddEndPuncttrue
\mciteSetBstMidEndSepPunct{\mcitedefaultmidpunct}
{\mcitedefaultendpunct}{\mcitedefaultseppunct}\relax
\EndOfBibitem
\bibitem[Umari \latin{et~al.}(2010)Umari, Stenuit, and Baroni]{umari_gw_2010}
Umari,~P.; Stenuit,~G.; Baroni,~S. GW quasiparticle spectra from occupied
  states only. \emph{Phys. Rev. B} \textbf{2010}, \emph{81}, 115104\relax
\mciteBstWouldAddEndPuncttrue
\mciteSetBstMidEndSepPunct{\mcitedefaultmidpunct}
{\mcitedefaultendpunct}{\mcitedefaultseppunct}\relax
\EndOfBibitem
\bibitem[Nguyen \latin{et~al.}(2012)Nguyen, Pham, Rocca, and Galli]{nguyen2012}
Nguyen,~H.-V.; Pham,~T.~A.; Rocca,~D.; Galli,~G. Improving Accuracy and
  Efficiency of Calculations of Photoemission Spectra within the Many-Body
  Perturbation Theory. \emph{Phys.~Rev.~B} \textbf{2012}, \emph{85},
  081101\relax
\mciteBstWouldAddEndPuncttrue
\mciteSetBstMidEndSepPunct{\mcitedefaultmidpunct}
{\mcitedefaultendpunct}{\mcitedefaultseppunct}\relax
\EndOfBibitem
\bibitem[Godby \latin{et~al.}(1988)Godby, Schl{\"u}ter, and Sham]{godby1988}
Godby,~R.~W.; Schl{\"u}ter,~M.; Sham,~L.~J. Self-Energy Operators and
  Exchange-Correlation Potentials in Semiconductors. \emph{Phys.~Rev.~B}
  \textbf{1988}, \emph{37}, 10159--10175\relax
\mciteBstWouldAddEndPuncttrue
\mciteSetBstMidEndSepPunct{\mcitedefaultmidpunct}
{\mcitedefaultendpunct}{\mcitedefaultseppunct}\relax
\EndOfBibitem
\bibitem[Giantomassi \latin{et~al.}(2011)Giantomassi, Stankovski, Shaltaf,
  Gr{\"u}ning, Bruneval, Rinke, and Rignanese]{giantomassi2011}
Giantomassi,~M.; Stankovski,~M.; Shaltaf,~R.; Gr{\"u}ning,~M.; Bruneval,~F.;
  Rinke,~P.; Rignanese,~G.-M. Electronic Properties of Interfaces and Defects
  from Many-Body Perturbation Theory: {{Recent}} Developments and Applications.
  \emph{Phys.~Status~Solidi~B} \textbf{2011}, \emph{248}, 275--289\relax
\mciteBstWouldAddEndPuncttrue
\mciteSetBstMidEndSepPunct{\mcitedefaultmidpunct}
{\mcitedefaultendpunct}{\mcitedefaultseppunct}\relax
\EndOfBibitem
\bibitem[van Schilfgaarde \latin{et~al.}(2006)van Schilfgaarde, Kotani, and
  Faleev]{vanschilfgaarde_quasiparticle_2006}
van Schilfgaarde,~M.; Kotani,~T.; Faleev,~S. Quasiparticle {Self}-{Consistent}
  {G} {W} {Theory}. \emph{Phys.~Rev.~Lett.~} \textbf{2006}, \emph{96},
  226402\relax
\mciteBstWouldAddEndPuncttrue
\mciteSetBstMidEndSepPunct{\mcitedefaultmidpunct}
{\mcitedefaultendpunct}{\mcitedefaultseppunct}\relax
\EndOfBibitem
\bibitem[Giannozzi \latin{et~al.}(2009)Giannozzi, Baroni, Bonini, Calandra,
  Car, Cavazzoni, Ceresoli, Chiarotti, Cococcioni, Dabo, Corso, de~Gironcoli,
  Fabris, Fratesi, Gebauer, Gerstmann, Gougoussis, Kokalj, Lazzeri,
  Martin-Samos, Marzari, Mauri, Mazzarello, Paolini, Pasquarello, Paulatto,
  Sbraccia, Scandolo, Sclauzero, Seitsonen, Smogunov, Umari, and
  Wentzcovitch]{Giannozzi2009}
Giannozzi,~P.; Baroni,~S.; Bonini,~N.; Calandra,~M.; Car,~R.; Cavazzoni,~C.;
  Ceresoli,~D.; Chiarotti,~G.~L.; Cococcioni,~M.; Dabo,~I.; Corso,~A.~D.;
  de~Gironcoli,~S.; Fabris,~S.; Fratesi,~G.; Gebauer,~R.; Gerstmann,~U.;
  Gougoussis,~C.; Kokalj,~A.; Lazzeri,~M.; Martin-Samos,~L.; Marzari,~N.;
  Mauri,~F.; Mazzarello,~R.; Paolini,~S.; Pasquarello,~A.; Paulatto,~L.;
  Sbraccia,~C.; Scandolo,~S.; Sclauzero,~G.; Seitsonen,~A.~P.; Smogunov,~A.;
  Umari,~P.; Wentzcovitch,~R.~M. {QUANTUM} {ESPRESSO}: a modular and
  open-source software project for quantum simulations of materials. \emph{J.
  Phys.: Condens. Matter} \textbf{2009}, \emph{21}, 395502\relax
\mciteBstWouldAddEndPuncttrue
\mciteSetBstMidEndSepPunct{\mcitedefaultmidpunct}
{\mcitedefaultendpunct}{\mcitedefaultseppunct}\relax
\EndOfBibitem
\bibitem[Perdew \latin{et~al.}(1996)Perdew, Burke, and Ernzerhof]{Perdew1996}
Perdew,~J.~P.; Burke,~K.; Ernzerhof,~M. Generalized Gradient Approximation Made
  Simple. \emph{Phys. Rev. Lett.} \textbf{1996}, \emph{77}, 3865--3868\relax
\mciteBstWouldAddEndPuncttrue
\mciteSetBstMidEndSepPunct{\mcitedefaultmidpunct}
{\mcitedefaultendpunct}{\mcitedefaultseppunct}\relax
\EndOfBibitem
\bibitem[Skone \latin{et~al.}(2014)Skone, Govoni, and Galli]{Skone2014}
Skone,~J.~H.; Govoni,~M.; Galli,~G. Self-consistent hybrid functional for
  condensed systems. \emph{Phys. Rev. B} \textbf{2014}, \emph{89}, 195112\relax
\mciteBstWouldAddEndPuncttrue
\mciteSetBstMidEndSepPunct{\mcitedefaultmidpunct}
{\mcitedefaultendpunct}{\mcitedefaultseppunct}\relax
\EndOfBibitem
\bibitem[Schlipf and Gygi(2015)Schlipf, and Gygi]{Schlipf2015}
Schlipf,~M.; Gygi,~F. Optimization algorithm for the generation of {ONCV}
  pseudopotentials. \emph{Comput. Phys. Commun.} \textbf{2015}, \emph{196},
  36--44\relax
\mciteBstWouldAddEndPuncttrue
\mciteSetBstMidEndSepPunct{\mcitedefaultmidpunct}
{\mcitedefaultendpunct}{\mcitedefaultseppunct}\relax
\EndOfBibitem
\bibitem[Sun \latin{et~al.}(2018)Sun, Berkelbach, Blunt, Booth, Guo, Li, Liu,
  McClain, Sayfutyarova, Sharma, Wouters, and Chan]{sun2018}
Sun,~Q.; Berkelbach,~T.~C.; Blunt,~N.~S.; Booth,~G.~H.; Guo,~S.; Li,~Z.;
  Liu,~J.; McClain,~J.~D.; Sayfutyarova,~E.~R.; Sharma,~S.; Wouters,~S.;
  Chan,~G. K.-L. {{PySCF}}: The {{Python}}-Based Simulations of Chemistry
  Framework. \emph{WIREs~Comput.~Mol.~Sci.} \textbf{2018}, \emph{8},
  e1340\relax
\mciteBstWouldAddEndPuncttrue
\mciteSetBstMidEndSepPunct{\mcitedefaultmidpunct}
{\mcitedefaultendpunct}{\mcitedefaultseppunct}\relax
\EndOfBibitem
\bibitem[Davies \latin{et~al.}(1976)Davies, Hamer, and Price]{Davies1976}
Davies,~G.; Hamer,~M.~F.; Price,~W.~C. Optical Studies of the 1.945 {{eV}}
  Vibronic Band in Diamond. \emph{Proc.~R.~Soc.~London~A} \textbf{1976},
  \emph{348}, 285--298\relax
\mciteBstWouldAddEndPuncttrue
\mciteSetBstMidEndSepPunct{\mcitedefaultmidpunct}
{\mcitedefaultendpunct}{\mcitedefaultseppunct}\relax
\EndOfBibitem
\bibitem[Rogers \latin{et~al.}(2008)Rogers, Armstrong, Sellars, and
  Manson]{Rogers2008}
Rogers,~L.~J.; Armstrong,~S.; Sellars,~M.~J.; Manson,~N.~B. Infrared emission
  of the {NV} centre in diamond: Zeeman and uniaxial stress studies. \emph{New
  J. Phys.} \textbf{2008}, \emph{10}, 103024\relax
\mciteBstWouldAddEndPuncttrue
\mciteSetBstMidEndSepPunct{\mcitedefaultmidpunct}
{\mcitedefaultendpunct}{\mcitedefaultseppunct}\relax
\EndOfBibitem
\bibitem[Goldman \latin{et~al.}(2015)Goldman, Sipahigil, Doherty, Yao, Bennett,
  Markham, Twitchen, Manson, Kubanek, and Lukin]{goldman2015}
Goldman,~M.~L.; Sipahigil,~A.; Doherty,~M.~W.; Yao,~N.~Y.; Bennett,~S.~D.;
  Markham,~M.; Twitchen,~D.~J.; Manson,~N.~B.; Kubanek,~A.; Lukin,~M.~D.
  Phonon-{{Induced Population Dynamics}} and {{Intersystem Crossing}} in
  {{Nitrogen-Vacancy Centers}}. \emph{Phys.~Rev.~Lett.~} \textbf{2015},
  \emph{114}, 145502\relax
\mciteBstWouldAddEndPuncttrue
\mciteSetBstMidEndSepPunct{\mcitedefaultmidpunct}
{\mcitedefaultendpunct}{\mcitedefaultseppunct}\relax
\EndOfBibitem
\bibitem[Doherty \latin{et~al.}(2011)Doherty, Manson, Delaney, and
  Hollenberg]{Doherty2011}
Doherty,~M.~W.; Manson,~N.~B.; Delaney,~P.; Hollenberg,~L. C.~L. The negatively
  charged nitrogen-vacancy centre in diamond: the electronic solution.
  \emph{New J. Phys.} \textbf{2011}, \emph{13}, 025019\relax
\mciteBstWouldAddEndPuncttrue
\mciteSetBstMidEndSepPunct{\mcitedefaultmidpunct}
{\mcitedefaultendpunct}{\mcitedefaultseppunct}\relax
\EndOfBibitem
\bibitem[Maze \latin{et~al.}(2011)Maze, Gali, Togan, Chu, Trifonov, Kaxiras,
  and Lukin]{Maze2011}
Maze,~J.~R.; Gali,~A.; Togan,~E.; Chu,~Y.; Trifonov,~A.; Kaxiras,~E.;
  Lukin,~M.~D. Properties of nitrogen-vacancy centers in diamond: the group
  theoretic approach. \emph{New J. Phys.} \textbf{2011}, \emph{13},
  025025\relax
\mciteBstWouldAddEndPuncttrue
\mciteSetBstMidEndSepPunct{\mcitedefaultmidpunct}
{\mcitedefaultendpunct}{\mcitedefaultseppunct}\relax
\EndOfBibitem
\bibitem[Choi \latin{et~al.}(2012)Choi, Jain, and Louie]{Choi2012}
Choi,~S.; Jain,~M.; Louie,~S.~G. Mechanism for optical initialization of spin
  in {NV}-center in diamond. \emph{Phys. Rev. B} \textbf{2012}, \emph{86},
  041202\relax
\mciteBstWouldAddEndPuncttrue
\mciteSetBstMidEndSepPunct{\mcitedefaultmidpunct}
{\mcitedefaultendpunct}{\mcitedefaultseppunct}\relax
\EndOfBibitem
\bibitem[Loubser and van Wyk(1978)Loubser, and van Wyk]{loubser1978}
Loubser,~J. H.~N.; van Wyk,~J.~A. Electron Spin Resonance in the Study of
  Diamond. \emph{Rep.~Prog.~Phys.~} \textbf{1978}, \emph{41}, 1201--1248\relax
\mciteBstWouldAddEndPuncttrue
\mciteSetBstMidEndSepPunct{\mcitedefaultmidpunct}
{\mcitedefaultendpunct}{\mcitedefaultseppunct}\relax
\EndOfBibitem
\bibitem[Doherty \latin{et~al.}(2013)Doherty, Manson, Delaney, Jelezko,
  Wrachtrup, and Hollenberg]{Doherty2013}
Doherty,~M.~W.; Manson,~N.~B.; Delaney,~P.; Jelezko,~F.; Wrachtrup,~J.;
  Hollenberg,~L.~C. The nitrogen-vacancy colour centre in diamond. \emph{Phys.
  Rep.} \textbf{2013}, \emph{528}, 1--45\relax
\mciteBstWouldAddEndPuncttrue
\mciteSetBstMidEndSepPunct{\mcitedefaultmidpunct}
{\mcitedefaultendpunct}{\mcitedefaultseppunct}\relax
\EndOfBibitem
\bibitem[Skone \latin{et~al.}(2016)Skone, Govoni, and Galli]{Skone2016}
Skone,~J.~H.; Govoni,~M.; Galli,~G. Nonempirical range-separated hybrid
  functionals for solids and molecules. \emph{Phys. Rev. B} \textbf{2016},
  \emph{93}, 235106\relax
\mciteBstWouldAddEndPuncttrue
\mciteSetBstMidEndSepPunct{\mcitedefaultmidpunct}
{\mcitedefaultendpunct}{\mcitedefaultseppunct}\relax
\EndOfBibitem
\bibitem[Brawand \latin{et~al.}(2016)Brawand, V\"or\"os, Govoni, and
  Galli]{Brawand2016}
Brawand,~N.~P.; V\"or\"os,~M.; Govoni,~M.; Galli,~G. Generalization of
  Dielectric-Dependent Hybrid Functionals to Finite Systems. \emph{Phys. Rev.
  X} \textbf{2016}, \emph{6}, 041002\relax
\mciteBstWouldAddEndPuncttrue
\mciteSetBstMidEndSepPunct{\mcitedefaultmidpunct}
{\mcitedefaultendpunct}{\mcitedefaultseppunct}\relax
\EndOfBibitem
\bibitem[Brawand \latin{et~al.}(2017)Brawand, Govoni, Vörös, and
  Galli]{Brawand2017}
Brawand,~N.~P.; Govoni,~M.; Vörös,~M.; Galli,~G. Performance and
  Self-Consistency of the Generalized Dielectric Dependent Hybrid Functional.
  \emph{J. Chem. Theory Comput.} \textbf{2017}, \emph{13}, 3318--3325, PMID:
  28537727\relax
\mciteBstWouldAddEndPuncttrue
\mciteSetBstMidEndSepPunct{\mcitedefaultmidpunct}
{\mcitedefaultendpunct}{\mcitedefaultseppunct}\relax
\EndOfBibitem
\bibitem[Gerosa \latin{et~al.}(2017)Gerosa, Bottani, Valentin, Onida, and
  Pacchioni]{Gerosa_2017}
Gerosa,~M.; Bottani,~C.~E.; Valentin,~C.~D.; Onida,~G.; Pacchioni,~G. Accuracy
  of dielectric-dependent hybrid functionals in the prediction of
  optoelectronic properties of metal oxide semiconductors: a comprehensive
  comparison with many-body GW and experiments. \emph{J. Phys.: Condens.
  Matter} \textbf{2017}, \emph{30}, 044003\relax
\mciteBstWouldAddEndPuncttrue
\mciteSetBstMidEndSepPunct{\mcitedefaultmidpunct}
{\mcitedefaultendpunct}{\mcitedefaultseppunct}\relax
\EndOfBibitem
\bibitem[Zheng \latin{et~al.}(2019)Zheng, Govoni, and Galli]{Zheng2019}
Zheng,~H.; Govoni,~M.; Galli,~G. Dielectric-dependent hybrid functionals for
  heterogeneous materials. \emph{Phys. Rev. Materials} \textbf{2019}, \emph{3},
  073803\relax
\mciteBstWouldAddEndPuncttrue
\mciteSetBstMidEndSepPunct{\mcitedefaultmidpunct}
{\mcitedefaultendpunct}{\mcitedefaultseppunct}\relax
\EndOfBibitem
\bibitem[Kehayias \latin{et~al.}(2013)Kehayias, Doherty, English, Fischer,
  Jarmola, Jensen, Leefer, Hemmer, Manson, and Budker]{kehayias2013}
Kehayias,~P.; Doherty,~M.~W.; English,~D.; Fischer,~R.; Jarmola,~A.;
  Jensen,~K.; Leefer,~N.; Hemmer,~P.; Manson,~N.~B.; Budker,~D. Infrared
  Absorption Band and Vibronic Structure of the Nitrogen-Vacancy Center in
  Diamond. \emph{Phys.~Rev.~B} \textbf{2013}, \emph{88}, 165202\relax
\mciteBstWouldAddEndPuncttrue
\mciteSetBstMidEndSepPunct{\mcitedefaultmidpunct}
{\mcitedefaultendpunct}{\mcitedefaultseppunct}\relax
\EndOfBibitem
\bibitem[Goldman \latin{et~al.}(2015)Goldman, Doherty, Sipahigil, Yao, Bennett,
  Manson, Kubanek, and Lukin]{goldman2015a}
Goldman,~M.~L.; Doherty,~M.~W.; Sipahigil,~A.; Yao,~N.~Y.; Bennett,~S.~D.;
  Manson,~N.~B.; Kubanek,~A.; Lukin,~M.~D. State-Selective Intersystem Crossing
  in Nitrogen-Vacancy Centers. \emph{Phys.~Rev.~B} \textbf{2015}, \emph{91},
  165201\relax
\mciteBstWouldAddEndPuncttrue
\mciteSetBstMidEndSepPunct{\mcitedefaultmidpunct}
{\mcitedefaultendpunct}{\mcitedefaultseppunct}\relax
\EndOfBibitem
\bibitem[Ma \latin{et~al.}(2010)Ma, Rohlfing, and Gali]{ma2010}
Ma,~Y.; Rohlfing,~M.; Gali,~A. Excited States of the Negatively Charged
  Nitrogen-Vacancy Color Center in Diamond. \emph{Phys.~Rev.~B} \textbf{2010},
  \emph{81}, 041204\relax
\mciteBstWouldAddEndPuncttrue
\mciteSetBstMidEndSepPunct{\mcitedefaultmidpunct}
{\mcitedefaultendpunct}{\mcitedefaultseppunct}\relax
\EndOfBibitem
\bibitem[Bhandari \latin{et~al.}(2021)Bhandari, Wysocki, Economou, Dev, and
  Park]{bhandari2021}
Bhandari,~C.; Wysocki,~A.~L.; Economou,~S.~E.; Dev,~P.; Park,~K.
  Multiconfigurational Study of the Negatively Charged Nitrogen-Vacancy Center
  in Diamond. \emph{Phys.~Rev.~B} \textbf{2021}, \emph{103}, 014115\relax
\mciteBstWouldAddEndPuncttrue
\mciteSetBstMidEndSepPunct{\mcitedefaultmidpunct}
{\mcitedefaultendpunct}{\mcitedefaultseppunct}\relax
\EndOfBibitem
\bibitem[Lin \latin{et~al.}(2008)Lin, Wang, Chang, Hayashi, and Lin]{lin2008}
Lin,~C.-K.; Wang,~Y.-H.; Chang,~H.-C.; Hayashi,~M.; Lin,~S.~H. One- and
  Two-Photon Absorption Properties of Diamond Nitrogen-Vacancy Defect Centers:
  {{A}} Theoretical Study. \emph{J.~Chem.~Phys.~} \textbf{2008}, \emph{129},
  124714\relax
\mciteBstWouldAddEndPuncttrue
\mciteSetBstMidEndSepPunct{\mcitedefaultmidpunct}
{\mcitedefaultendpunct}{\mcitedefaultseppunct}\relax
\EndOfBibitem
\bibitem[Zyubin \latin{et~al.}(2009)Zyubin, Mebel, Hayashi, Chang, and
  Lin]{zyubin2009}
Zyubin,~A.~S.; Mebel,~A.~M.; Hayashi,~M.; Chang,~H.~C.; Lin,~S.~H. Quantum
  Chemical Modeling of Photoadsorption Properties of the Nitrogen-Vacancy Point
  Defect in Diamond. \emph{J.~Comput.~Chem.~} \textbf{2009}, \emph{30},
  119--131\relax
\mciteBstWouldAddEndPuncttrue
\mciteSetBstMidEndSepPunct{\mcitedefaultmidpunct}
{\mcitedefaultendpunct}{\mcitedefaultseppunct}\relax
\EndOfBibitem
\bibitem[Delaney \latin{et~al.}(2010)Delaney, Greer, and Larsson]{delaney2010}
Delaney,~P.; Greer,~J.~C.; Larsson,~J.~A. Spin-{{Polarization Mechanisms}} of
  the {{Nitrogen-Vacancy Center}} in {{Diamond}}. \emph{Nano~Lett.~}
  \textbf{2010}, \emph{10}, 610--614\relax
\mciteBstWouldAddEndPuncttrue
\mciteSetBstMidEndSepPunct{\mcitedefaultmidpunct}
{\mcitedefaultendpunct}{\mcitedefaultseppunct}\relax
\EndOfBibitem
\bibitem[Gali and Maze(2013)Gali, and Maze]{Gali2013}
Gali,~A.; Maze,~J.~R. Ab initio study of the split silicon-vacancy defect in
  diamond: Electronic structure and related properties. \emph{Phys. Rev. B}
  \textbf{2013}, \emph{88}, 235205\relax
\mciteBstWouldAddEndPuncttrue
\mciteSetBstMidEndSepPunct{\mcitedefaultmidpunct}
{\mcitedefaultendpunct}{\mcitedefaultseppunct}\relax
\EndOfBibitem
\bibitem[Thiering and Gali(2018)Thiering, and Gali]{Thiering2018}
Thiering,~G. m.~H.; Gali,~A. Ab Initio Magneto-Optical Spectrum of Group-IV
  Vacancy Color Centers in Diamond. \emph{Phys. Rev. X} \textbf{2018},
  \emph{8}, 021063\relax
\mciteBstWouldAddEndPuncttrue
\mciteSetBstMidEndSepPunct{\mcitedefaultmidpunct}
{\mcitedefaultendpunct}{\mcitedefaultseppunct}\relax
\EndOfBibitem
\bibitem[Green \latin{et~al.}(2019)Green, Doherty, Nako, Manson,
  {D'Haenens-Johansson}, Williams, Twitchen, and Newton]{Green2019}
Green,~B.~L.; Doherty,~M.~W.; Nako,~E.; Manson,~N.~B.;
  {D'Haenens-Johansson},~U. F.~S.; Williams,~S.~D.; Twitchen,~D.~J.;
  Newton,~M.~E. Electronic Structure of the Neutral Silicon-Vacancy Center in
  Diamond. \emph{Phys.~Rev.~B} \textbf{2019}, \emph{99}, 161112\relax
\mciteBstWouldAddEndPuncttrue
\mciteSetBstMidEndSepPunct{\mcitedefaultmidpunct}
{\mcitedefaultendpunct}{\mcitedefaultseppunct}\relax
\EndOfBibitem
\bibitem[Thiering and Gali(2019)Thiering, and Gali]{Thiering2019}
Thiering,~G.; Gali,~A. The (eg $\otimes$ eu) $\otimes$ Eg product
  Jahn{\textendash}Teller effect in the neutral group-{IV} vacancy quantum bits
  in diamond. \emph{npj Comput. Mater.} \textbf{2019}, \emph{5}, 18\relax
\mciteBstWouldAddEndPuncttrue
\mciteSetBstMidEndSepPunct{\mcitedefaultmidpunct}
{\mcitedefaultendpunct}{\mcitedefaultseppunct}\relax
\EndOfBibitem
\bibitem[Zhang \latin{et~al.}(2020)Zhang, Stevenson, Thiering, Rose, Huang,
  Edmonds, Markham, Lyon, Gali, and {de Leon}]{zhang2020a}
Zhang,~Z.-H.; Stevenson,~P.; Thiering,~G.; Rose,~B.~C.; Huang,~D.;
  Edmonds,~A.~M.; Markham,~M.~L.; Lyon,~S.~A.; Gali,~A.; {de Leon},~N.~P.
  Optically {{Detected Magnetic Resonance}} in {{Neutral Silicon Vacancy
  Centers}} in {{Diamond}} via {{Bound Exciton States}}.
  \emph{Phys.~Rev.~Lett.~} \textbf{2020}, \emph{125}, 237402\relax
\mciteBstWouldAddEndPuncttrue
\mciteSetBstMidEndSepPunct{\mcitedefaultmidpunct}
{\mcitedefaultendpunct}{\mcitedefaultseppunct}\relax
\EndOfBibitem
\end{mcitethebibliography}


\providecommand{\latin}[1]{#1}
\makeatletter
\providecommand{\doi}
  {\begingroup\let\do\@makeother\dospecials
  \catcode`\{=1 \catcode`\}=2 \doi@aux}
\providecommand{\doi@aux}[1]{\endgroup\texttt{#1}}
\makeatother
\providecommand*\mcitethebibliography{\thebibliography}
\csname @ifundefined\endcsname{endmcitethebibliography}
  {\let\endmcitethebibliography\endthebibliography}{}
\begin{mcitethebibliography}{24}
\providecommand*\natexlab[1]{#1}
\providecommand*\mciteSetBstSublistMode[1]{}
\providecommand*\mciteSetBstMaxWidthForm[2]{}
\providecommand*\mciteBstWouldAddEndPuncttrue
  {\def\EndOfBibitem{\unskip.}}
\providecommand*\mciteBstWouldAddEndPunctfalse
  {\let\EndOfBibitem\relax}
\providecommand*\mciteSetBstMidEndSepPunct[3]{}
\providecommand*\mciteSetBstSublistLabelBeginEnd[3]{}
\providecommand*\EndOfBibitem{}
\mciteSetBstSublistMode{f}
\mciteSetBstMaxWidthForm{subitem}{(\alph{mcitesubitemcount})}
\mciteSetBstSublistLabelBeginEnd
  {\mcitemaxwidthsubitemform\space}
  {\relax}
  {\relax}

\bibitem[Govoni and Galli(2015)Govoni, and Galli]{govoni2015}
Govoni,~M.; Galli,~G. Large Scale {{GW}} Calculations.
  \emph{J.~Chem.~Theory.~Comput.~} \textbf{2015}, \emph{11}, 2680--2696\relax
\mciteBstWouldAddEndPuncttrue
\mciteSetBstMidEndSepPunct{\mcitedefaultmidpunct}
{\mcitedefaultendpunct}{\mcitedefaultseppunct}\relax
\EndOfBibitem
\bibitem[Ma \latin{et~al.}(2020)Ma, Sheng, Govoni, and Galli]{ma2020}
Ma,~H.; Sheng,~N.; Govoni,~M.; Galli,~G. First-Principles Studies of Strongly
  Correlated States in Defect Spin Qubits in Diamond.
  \emph{Phys.~Chem.~Chem.~Phys.~} \textbf{2020}, \emph{22}, 25522--25527\relax
\mciteBstWouldAddEndPuncttrue
\mciteSetBstMidEndSepPunct{\mcitedefaultmidpunct}
{\mcitedefaultendpunct}{\mcitedefaultseppunct}\relax
\EndOfBibitem
\bibitem[Ma \latin{et~al.}(2020)Ma, Govoni, and Galli]{ma2020a}
Ma,~H.; Govoni,~M.; Galli,~G. Quantum Simulations of Materials on Near-Term
  Quantum Computers. \emph{npj~Comput.~Mater.} \textbf{2020}, \emph{6},
  1--8\relax
\mciteBstWouldAddEndPuncttrue
\mciteSetBstMidEndSepPunct{\mcitedefaultmidpunct}
{\mcitedefaultendpunct}{\mcitedefaultseppunct}\relax
\EndOfBibitem
\bibitem[Ma \latin{et~al.}(2021)Ma, Sheng, Govoni, and Galli]{ma2021}
Ma,~H.; Sheng,~N.; Govoni,~M.; Galli,~G. Quantum Embedding Theory for Strongly
  Correlated States in Materials. \emph{J.~Chem.~Theory.~Comput.~}
  \textbf{2021}, \emph{17}, 2116--2125\relax
\mciteBstWouldAddEndPuncttrue
\mciteSetBstMidEndSepPunct{\mcitedefaultmidpunct}
{\mcitedefaultendpunct}{\mcitedefaultseppunct}\relax
\EndOfBibitem
\bibitem[Muechler \latin{et~al.}(2021)Muechler, Badrtdinov, Hampel, Cano,
  Rösner, and Dreyer]{muechler_quantum_2021}
Muechler,~L.; Badrtdinov,~D.~I.; Hampel,~A.; Cano,~J.; Rösner,~M.;
  Dreyer,~C.~E. Quantum embedding methods for correlated excited states of
  point defects: {Case} studies and challenges. \emph{arXiv:2105.08705
  [cond-mat]} \textbf{2021}, arXiv: 2105.08705\relax
\mciteBstWouldAddEndPuncttrue
\mciteSetBstMidEndSepPunct{\mcitedefaultmidpunct}
{\mcitedefaultendpunct}{\mcitedefaultseppunct}\relax
\EndOfBibitem
\bibitem[Georges and Kotliar(1992)Georges, and Kotliar]{georges1992}
Georges,~A.; Kotliar,~G. Hubbard Model in Infinite Dimensions.
  \emph{Phys.~Rev.~B} \textbf{1992}, \emph{45}, 6479--6483\relax
\mciteBstWouldAddEndPuncttrue
\mciteSetBstMidEndSepPunct{\mcitedefaultmidpunct}
{\mcitedefaultendpunct}{\mcitedefaultseppunct}\relax
\EndOfBibitem
\bibitem[Georges \latin{et~al.}(1996)Georges, Kotliar, Krauth, and
  Rozenberg]{georges1996}
Georges,~A.; Kotliar,~G.; Krauth,~W.; Rozenberg,~M.~J. Dynamical Mean-Field
  Theory of Strongly Correlated Fermion Systems and the Limit of Infinite
  Dimensions. \emph{Rev.~Mod.~Phys.~} \textbf{1996}, \emph{68}, 13--125\relax
\mciteBstWouldAddEndPuncttrue
\mciteSetBstMidEndSepPunct{\mcitedefaultmidpunct}
{\mcitedefaultendpunct}{\mcitedefaultseppunct}\relax
\EndOfBibitem
\bibitem[Georges(2004)]{georges2004}
Georges,~A. Strongly Correlated Electron Materials: Dynamical Mean-Field Theory
  and Electronic Structure. \emph{AIP~Conf.~Proc.~} \textbf{2004}, \emph{715},
  3--74\relax
\mciteBstWouldAddEndPuncttrue
\mciteSetBstMidEndSepPunct{\mcitedefaultmidpunct}
{\mcitedefaultendpunct}{\mcitedefaultseppunct}\relax
\EndOfBibitem
\bibitem[Anisimov \latin{et~al.}(1997)Anisimov, Poteryaev, Korotin, Anokhin,
  and Kotliar]{anisimov1997}
Anisimov,~V.~I.; Poteryaev,~A.~I.; Korotin,~M.~A.; Anokhin,~A.~O.; Kotliar,~G.
  First-Principles Calculations of the Electronic Structure and Spectra of
  Strongly Correlated Systems: Dynamical Mean-Field Theory.
  \emph{J.~Phys.~Condens.~Matter.~} \textbf{1997}, \emph{9}, 7359--7367\relax
\mciteBstWouldAddEndPuncttrue
\mciteSetBstMidEndSepPunct{\mcitedefaultmidpunct}
{\mcitedefaultendpunct}{\mcitedefaultseppunct}\relax
\EndOfBibitem
\bibitem[Kotliar \latin{et~al.}(2006)Kotliar, Savrasov, Haule, Oudovenko,
  Parcollet, and Marianetti]{kotliar2006}
Kotliar,~G.; Savrasov,~S.~Y.; Haule,~K.; Oudovenko,~V.~S.; Parcollet,~O.;
  Marianetti,~C.~A. Electronic Structure Calculations with Dynamical Mean-Field
  Theory. \emph{Rev.~Mod.~Phys.~} \textbf{2006}, \emph{78}, 865--951\relax
\mciteBstWouldAddEndPuncttrue
\mciteSetBstMidEndSepPunct{\mcitedefaultmidpunct}
{\mcitedefaultendpunct}{\mcitedefaultseppunct}\relax
\EndOfBibitem
\bibitem[Biermann \latin{et~al.}(2003)Biermann, Aryasetiawan, and
  Georges]{biermann2003}
Biermann,~S.; Aryasetiawan,~F.; Georges,~A. First-Principles Approach to the
  Electronic Structure of Strongly Correlated Systems: Combining the {{GW}}
  Approximation and Dynamical Mean-Field Theory. \emph{Phys.~Rev.~Lett.~}
  \textbf{2003}, \emph{90}, 086402\relax
\mciteBstWouldAddEndPuncttrue
\mciteSetBstMidEndSepPunct{\mcitedefaultmidpunct}
{\mcitedefaultendpunct}{\mcitedefaultseppunct}\relax
\EndOfBibitem
\bibitem[Biermann(2014)]{biermann2014}
Biermann,~S. Dynamical Screening Effects in Correlated Electron
  Materials\textemdash a Progress Report on Combined Many-Body Perturbation and
  Dynamical Mean Field Theory: '{{GW+DMFT}}'. \emph{J.~Phys.~Condens.~Matter.~}
  \textbf{2014}, \emph{26}, 173202\relax
\mciteBstWouldAddEndPuncttrue
\mciteSetBstMidEndSepPunct{\mcitedefaultmidpunct}
{\mcitedefaultendpunct}{\mcitedefaultseppunct}\relax
\EndOfBibitem
\bibitem[Boehnke \latin{et~al.}(2016)Boehnke, Nilsson, Aryasetiawan, and
  Werner]{boehnke2016}
Boehnke,~L.; Nilsson,~F.; Aryasetiawan,~F.; Werner,~P. When Strong Correlations
  Become Weak: Consistent Merging of {{GW}} and {{DMFT}}. \emph{Phys.~Rev.~B}
  \textbf{2016}, \emph{94}, 201106\relax
\mciteBstWouldAddEndPuncttrue
\mciteSetBstMidEndSepPunct{\mcitedefaultmidpunct}
{\mcitedefaultendpunct}{\mcitedefaultseppunct}\relax
\EndOfBibitem
\bibitem[Choi \latin{et~al.}(2016)Choi, Kutepov, Haule, {van Schilfgaarde}, and
  Kotliar]{choi2016}
Choi,~S.; Kutepov,~A.; Haule,~K.; {van Schilfgaarde},~M.; Kotliar,~G.
  First-Principles Treatment of {{Mott}} Insulators: Linearized {{QSGW+DMFT}}
  Approach. \emph{npj~Quantum~Mater.} \textbf{2016}, \emph{1}, 1--6\relax
\mciteBstWouldAddEndPuncttrue
\mciteSetBstMidEndSepPunct{\mcitedefaultmidpunct}
{\mcitedefaultendpunct}{\mcitedefaultseppunct}\relax
\EndOfBibitem
\bibitem[Nilsson \latin{et~al.}(2017)Nilsson, Boehnke, Werner, and
  Aryasetiawan]{nilsson2017}
Nilsson,~F.; Boehnke,~L.; Werner,~P.; Aryasetiawan,~F. Multitier
  Self-Consistent {{GW+EDMFT}}. \emph{Phys.~Rev.~Materials~} \textbf{2017},
  \emph{1}, 043803\relax
\mciteBstWouldAddEndPuncttrue
\mciteSetBstMidEndSepPunct{\mcitedefaultmidpunct}
{\mcitedefaultendpunct}{\mcitedefaultseppunct}\relax
\EndOfBibitem
\bibitem[Sun and Kotliar(2002)Sun, and Kotliar]{sun2002}
Sun,~P.; Kotliar,~G. Extended Dynamical Mean-Field Theory and {{GW}} Method.
  \emph{Phys.~Rev.~B} \textbf{2002}, \emph{66}, 085120\relax
\mciteBstWouldAddEndPuncttrue
\mciteSetBstMidEndSepPunct{\mcitedefaultmidpunct}
{\mcitedefaultendpunct}{\mcitedefaultseppunct}\relax
\EndOfBibitem
\bibitem[Tomczak \latin{et~al.}(2012)Tomczak, Casula, Miyake, Aryasetiawan, and
  Biermann]{tomczak2012}
Tomczak,~J.~M.; Casula,~M.; Miyake,~T.; Aryasetiawan,~F.; Biermann,~S. Combined
  {{GW}} and Dynamical Mean-Field Theory: {{Dynamical}} Screening Effects in
  Transition Metal Oxides. \emph{Europhys.~Lett.~} \textbf{2012}, \emph{100},
  67001\relax
\mciteBstWouldAddEndPuncttrue
\mciteSetBstMidEndSepPunct{\mcitedefaultmidpunct}
{\mcitedefaultendpunct}{\mcitedefaultseppunct}\relax
\EndOfBibitem
\bibitem[Lee and Haule(2017)Lee, and Haule]{lee2017}
Lee,~J.; Haule,~K. Diatomic Molecule as a Testbed for Combining {{DMFT}} with
  Electronic Structure Methods Such as {{GW}} and {{DFT}}. \emph{Phys.~Rev.~B}
  \textbf{2017}, \emph{95}, 155104\relax
\mciteBstWouldAddEndPuncttrue
\mciteSetBstMidEndSepPunct{\mcitedefaultmidpunct}
{\mcitedefaultendpunct}{\mcitedefaultseppunct}\relax
\EndOfBibitem
\bibitem[Smith and Si(2000)Smith, and Si]{smith2000}
Smith,~J.~L.; Si,~Q. Spatial Correlations in Dynamical Mean-Field Theory.
  \emph{Phys.~Rev.~B} \textbf{2000}, \emph{61}, 5184--5193\relax
\mciteBstWouldAddEndPuncttrue
\mciteSetBstMidEndSepPunct{\mcitedefaultmidpunct}
{\mcitedefaultendpunct}{\mcitedefaultseppunct}\relax
\EndOfBibitem
\bibitem[Wilson \latin{et~al.}(2008)Wilson, Gygi, and
  Galli]{wilson_efficient_2008}
Wilson,~H.~F.; Gygi,~F.; Galli,~G. Efficient iterative method for calculations
  of dielectric matrices. \emph{Phys.~Rev.~B} \textbf{2008}, \emph{78},
  113303\relax
\mciteBstWouldAddEndPuncttrue
\mciteSetBstMidEndSepPunct{\mcitedefaultmidpunct}
{\mcitedefaultendpunct}{\mcitedefaultseppunct}\relax
\EndOfBibitem
\bibitem[Thiering and Gali(2019)Thiering, and Gali]{Thiering2019}
Thiering,~G.; Gali,~A. The (eg $\otimes$ eu) $\otimes$ Eg product
  Jahn{\textendash}Teller effect in the neutral group-{IV} vacancy quantum bits
  in diamond. \emph{npj Comput. Mater.} \textbf{2019}, \emph{5}, 18\relax
\mciteBstWouldAddEndPuncttrue
\mciteSetBstMidEndSepPunct{\mcitedefaultmidpunct}
{\mcitedefaultendpunct}{\mcitedefaultseppunct}\relax
\EndOfBibitem
\bibitem[Davies \latin{et~al.}(1976)Davies, Hamer, and Price]{Davies1976}
Davies,~G.; Hamer,~M.~F.; Price,~W.~C. Optical Studies of the 1.945 {{eV}}
  Vibronic Band in Diamond. \emph{Proc.~R.~Soc.~London~A} \textbf{1976},
  \emph{348}, 285--298\relax
\mciteBstWouldAddEndPuncttrue
\mciteSetBstMidEndSepPunct{\mcitedefaultmidpunct}
{\mcitedefaultendpunct}{\mcitedefaultseppunct}\relax
\EndOfBibitem
\bibitem[Green \latin{et~al.}(2019)Green, Doherty, Nako, Manson,
  {D'Haenens-Johansson}, Williams, Twitchen, and Newton]{Green2019}
Green,~B.~L.; Doherty,~M.~W.; Nako,~E.; Manson,~N.~B.;
  {D'Haenens-Johansson},~U. F.~S.; Williams,~S.~D.; Twitchen,~D.~J.;
  Newton,~M.~E. Electronic Structure of the Neutral Silicon-Vacancy Center in
  Diamond. \emph{Phys.~Rev.~B} \textbf{2019}, \emph{99}, 161112\relax
\mciteBstWouldAddEndPuncttrue
\mciteSetBstMidEndSepPunct{\mcitedefaultmidpunct}
{\mcitedefaultendpunct}{\mcitedefaultseppunct}\relax
\EndOfBibitem
\end{mcitethebibliography}

\end{document}


\maketitle
\section{Definition of matrix elements}
In the main text, matrix elements are given in a short-hand notation. Here, we
provide detailed expressions of these matrix elements.

Matrix elements of the Hartree energy are given by
\begin{equation}\label{eq:matrix_h}
    \left[ V_{\mathrm{H}} \right]_{ij} 
    = \left[ v \rho \right]_{ij} = \sum_{kl} v_{ikjl} \rho_{kl},
\end{equation}
where the matrix elements of the bare Coulomb potential $v$, $v_{ikjl}$, are
defined in Eq.~2 of the main text. Analogously, the matrix elements of the
Hartree contribution to the double counting term in Eq.~25 and 26 of the
main text are given by
\begin{equation}\label{eq:matrix_xc}
    \left[ W^R_0 (\omega =0) \rho^A \right]_{ij} 
    = \sum_{kl}^A \left[ W^R_0(\omega = 0)\right]_{ikjl} \rho^A_{kl}.
\end{equation}
Here, $\rho_{kl}$ are the matrix elements of the density matrix. The matrix
elements of the $G_0W_0$-contribution to the self-energy in
Eq.~26 are defined as~\cite{govoni2015}
\begin{equation}
 \left[ \mathrm{i}G^R_0 W_0 \right]_{ij}(\omega) 
  = \mathrm{i} \langle \zeta_i |\int \frac{d\omega'}{2\pi} 
  G^R_0(\mathbf{r}, \mathbf{r}'; \omega +\omega')
  W_0(\mathbf{r}, \mathbf{r}'; \omega')| \zeta_j \rangle,
\end{equation}
where $\zeta_i$ are the single-particle functions that span
the active space. The reduced Kohn-Sham Green's function $G_0^R$ is defined in
Eq. 23 of the main text, the screened Coulomb potential $W_0$ in Eq. 19.

\clearpage

\section{Self-energy of the effective Hamiltonian} \label{sec:dc_self_energy}
In this section, we derive the Hartree, Hartree-Fock, and $G_0W_0$ self-energies for the effective Hamiltonian.
The effective Hamiltonian is given by
\begin{equation}
  H^{\mathrm{eff}} = \sum_{ij}^A t^{\mathrm{eff}}_{ij}a^\dagger_i a_j +
  \frac{1}{2} \sum_{ijkl}^A v^{\mathrm{eff}}_{ijkl}
  a^\dagger_i a^\dagger_j a_l a_k,
\end{equation}
where $t^{\mathrm{eff}}$ and $v^{\mathrm{eff}}$ are the effective one- and two-body terms, respectively. In the following, we assume that the active space is formed by a set of Kohn-Sham eigenstates. In this case, the independent-particle Green's function associated to the effective Hamiltonian $G^{\mathrm{eff}}_0$ is given by $G_0^A$. The density matrix $\rho^{\mathrm{eff}}$ associated to the effective Hamiltonian is given by $\rho^A$, which is the Kohn-Sham density matrix projected onto $A$.

The Hartree self-energy $\Sigma^{\mathrm{eff}}_{\mathrm{H}}$ of the effective Hamiltonian is given by
\begin{equation}
    \Sigma^{\mathrm{eff}}_{\mathrm{H}} = v^{\mathrm{eff}} \rho^{\mathrm{eff}} = v^{\mathrm{eff}}\rho^A .
\end{equation}
We note that $v^{\mathrm{eff}}$ is a renormalized Coulomb interaction that contains the effect of the screening generated by the electrons in the environment, but not the screening generated by the electrons in the active space.
\subsection{Hartree-Fock}
In the Hartree-Fock (HF) approximation, the polarizability is assumed to vanish, \textit{i.e.} $P_0^{\mathrm{eff}} \approx 0$, such that the interaction is is not screened by the electrons that belong to the active space, \textit{i.e.} $W_0^{\mathrm{eff}} = v^{\mathrm{eff}}$. Consequently, the Hartree-Fock self-energy of the effective Hamiltonian is given by
\begin{equation}\label{eq:eff_hf}
    \Sigma^{\mathrm{eff}}_{\mathrm{HF}} = \Sigma^{\mathrm{eff}}_{\mathrm{H}} + \Sigma^{\mathrm{eff}}_{\mathrm{x}} = \Sigma^{\mathrm{eff}}_{\mathrm{H}} + \mathrm{i}G_0^{\mathrm{eff}}v^{\mathrm{eff}} = v^{\mathrm{eff}} \rho^A + \mathrm{i}G_0^{A}v^{\mathrm{eff}},
\end{equation}
where we employ the matrix notations of Eqs.~\ref{eq:matrix_h} and \ref{eq:matrix_xc}.
\subsection{\texorpdfstring{$G_0W_0$}{}}
In the $G_0W_0$ approximation, the screening generated by the electrons in the active space is described the effective polarizability $P^{\mathrm{eff}}$. Within the the random-phase approximation, $P^{\mathrm{eff}} \approx P_0^{\mathrm{eff}} = P_0^A$. The screened potential is therefore $W_0^{\mathrm{eff}} = {\left[ {\left[v^{\mathrm{eff}}\right]}^{-1} - P_0^A \right]}^{-1}$, which includes the screening generated by the electrons in the environment (included in $v^{\mathrm{eff}}$) and the screening generated by the electrons in the active space (included in $P_0^A$). As a result, the self-energy in the $G_0W_0$ approximation is given by
\begin{equation}\label{eq:eff_gw}
    \Sigma^{\mathrm{eff}}_{G_0W_0} = \Sigma^{\mathrm{eff}}_{\mathrm{H}} + \mathrm{i}G_0^{\mathrm{eff}}W^{\mathrm{eff}}_0 = v^{\mathrm{eff}} \rho^A + \mathrm{i}G_0^AW_0^{\mathrm{eff}}.
\end{equation}

\section{Exact double counting at Hartree-Fock level of theory (EDC@HF)}
In the following, we derive a double counting term for the embedding of an active space into an environment that is described within the Hartree-Fock (HF) level of theory. Our derivation is analogous to the case of embedding in DFT+$G_0W_0$ used in the main manuscript.
To derive the
double counting terms $\Sigma^{\mathrm{dc}}$ and $P^{\mathrm{dc}}$ in this case,
we again require that the chain rule be satisfied, \textit{i.e.}, that the
self-energy and polarizability  of the whole system ($A+E$), evaluated at the
low-level of theory, be identical to that of $A$ embedded in $E$. Within the HF
approximation this requirement yields 
\begin{align}\label{eq:dc_hf}
  &\Sigma^{\mathrm{dc}} = \Sigma_{\mathrm{HF}}^{\mathrm{eff}}  \\
  &P^{\mathrm{dc}} = P_0^{\mathrm{eff}}.
\end{align}
Within HF we have that $P_0=0$ and $W^R_0=v$.  Hence, the effective interaction in the active space is given by the bare Coulomb potential, \textit{i.e.} $v^{\mathrm{eff}} = v$.

The double counting contribution to the self-energy, $\Sigma^{\mathrm{dc}}$ is obtained by inserting the HF self-energy of the effective Hamiltonian (Eq.~\ref{eq:eff_hf}) into the double-counting expression in Eq.~\ref{eq:dc_hf}. We thus obtain
\begin{equation}
  \Sigma^{\mathrm{dc}} = v \rho^A + \mathrm{i}G_0^A  v.
\end{equation}
Having obtained  explicit expressions for the double counting terms, we can finally
determine the one-body parameters of the effective Hamiltonian. We write $G^R$ as
\begin{equation}
  \begin{aligned}
    \left[ G^R \right]^{-1} &= g^{-1} - \left[ V_{\mathrm{H}} + \mathrm{i}G_0v
    \right] +
    \left[ v  \rho^A + \mathrm{i}G_0^A v \right]\\ 
    &= \omega - H^{\mathrm{HF}} + v \rho^A + \mathrm{i}G_0^A v
  \end{aligned}
\end{equation}
From the equation above, we obtain the double counting contribution to the
effective one-body terms in the HF scheme:
\begin{equation}
  t^{\mathrm{dc}} = v \rho^A  + \mathrm{i}G_0^A v.
\end{equation}
By definition, this HF double counting correction — when applied to HF —
satisfies the chain rule and hence it does not introduce errors that originate
from the separation of the system into active space and the environment.

However, the HF double counting (HFDC) term used in
Ref.~\citenum{ma2020,ma2020a,ma2021,muechler_quantum_2021}, and reported in
Eq.~6 of the main manuscript, contains two inconsistencies. The
first inconsistency is given by the fact that the double-counting term is
inspired by HF but is applied to DFT, hence leading to an approximate
cancellation of double counting effects. Additionally, the
polarizability is computed at the RPA level of theory and thus a screened
Coulomb interaction instead of the bare Coulomb interaction is used. Instead,
the double counting scheme discussed in the main text can be applied to DFT and
consistently applies $G_0W_0$ to the mean-field electronic structure. 

\clearpage

\section{Comparison with other double counting schemes in the literature}
In this section, we compare the double counting (DC) scheme introduced in the
main text to schemes adopted in other Green's function embedding theories
including dynamical mean-field theory. To simplify the comparison, we consider
the case of an active space composed of a set of localized correlated orbitals
within the environment of a solid or condensed system whose Brillouin zone (BZ)
is sampled by the $\Gamma$-point only. In dynamical mean-field theory
(DMFT)~\cite{georges1992,georges1996,georges2004,anisimov1997,kotliar2006},
\textit{local} quantities are defined as an average over $\mathbf{k}$-point,
\textit{e.g.}
$\Sigma^{\mathrm{loc}}=\frac{1}{N_{\mathbf{k}}}\sum_{\mathbf{k}}\Sigma(\mathbf{k})$
and the \textit{non-local} quantities as the remainders, \textit{e.g.}
$\Sigma^{\mathrm{non-loc}}(\mathbf{k}) = \Sigma(\mathbf{k}) -
\Sigma^{\mathrm{loc}}$. In supercells with $\Gamma$-point sampling, every
quantity is local by definition (all non-local contributions vanish).

\subsection{Dynamical mean-field theory (DMFT)}
We denote with
DMFT+$GW$~\cite{biermann2003,biermann2014,boehnke2016,choi2016,nilsson2017,sun2002}
a group of quantum-embedding methodologies that combine dynamical mean-field
theory (DMFT) to describe the active space with the $GW$ approximation for the
environment. In this case, the self energy is expressed as~\cite{tomczak2012}
\begin{equation}
    \Sigma = \Sigma^{\mathrm{DMFT}} + \Sigma^{GW} - \Sigma^{\mathrm{dc}},
\end{equation}
where the double counting contribution $\Sigma^{\mathrm{dc}}$ is given
by~\cite{tomczak2012,lee2017}
\begin{equation}\label{eq:dc1}
    \Sigma^{\mathrm{dc}} = G^{\mathrm{loc},A}W^{\mathrm{loc}}.
\end{equation}
As mentioned above, when the Brillouin zone of the supercell is sampled solely
with the $\Gamma$-point,  all quantitites are local. Hence, using $G^{A} =
f^{A}G f^{A}$, it is easy to show that the double counting adopted within
DMFT+$GW$ is identical to the one derived in Eq. 26 in the main text.

We note that a double counting correction for DMFT+$GW$ different to the one in
Eq.~\ref{eq:dc1}, $\Sigma^{\mathrm{dc2}}$, has been suggested in
Refs.~\cite{tomczak2012,lee2017}, given by
\begin{equation}
    \Sigma^{\mathrm{dc2}} 
    = \Sigma^{\mathrm{loc},A} \neq G^{\mathrm{loc},A}W^{\mathrm{loc}}.
\end{equation}
In this case, the projection on the subspace $A$ is performed after the
calculation of the self energy, and $\Sigma^{\mathrm{dc2}}$ contains
contribution from the environment as well. Using the definition of $\Sigma$ in
the $GW$ approximation,
\begin{equation}
    \Sigma_{kl}(\tau) = - \sum_{mn}^{A+E} G_{mn}(\tau)W_{mknl}(\tau),
\end{equation}
where $\tau$ is the imaginary time, $\Sigma^{\mathrm{dc2}}$
is given by
\begin{equation}\label{eq:dmft-cd2}
  \begin{aligned}
    \Sigma^{\mathrm{dc2}}_{kl} &= f^{A} \sum^{A+ E}_{mn} G_{mn}W_{mknl}f^{A}\\
    &= f^{A} \sum_{m}^{A} G_{mm}W_{mkml} f^{A} + f^{A} \sum_{m}^{E} G_{mm}W_{mkml}f^{A}\\
    &= \Sigma^{\mathrm{dc}} + f^{A} \sum_{m}^{E} G_{mm}W_{mkml}f^{A},
  \end{aligned}
\end{equation}
where we took into account that the Green's function is diagonal in the
single-particle space within the quasiparticle approximation.
Equation~\ref{eq:dmft-cd2} shows that $\Sigma^{\mathrm{dc2}}$ differs from
$\Sigma^{\mathrm{dc}}$ by an additional term with an explicit summation over
states in the environment $E$. While Ref.~\cite{tomczak2012} initially found
that there was no numerical difference between $\Sigma^{\mathrm{dc}}$ and
$\Sigma^{\mathrm{dc2}}$ in DMFT+$GW$ calculations, Ref.~\cite{lee2017} later
showed that $\Sigma^{\mathrm{dc2}}$ corrects non-causal contributions in the
embedding of DMFT and quasiparticle self-consistent $GW$ (DMFT+QS$GW$). We note
that $\Sigma^{\mathrm{dc2}}$ violates the chain rule.

\subsection{Extended dynamical mean-field theory (EDMFT)}
While in DMFT the interaction in the active space is
described by a set of Hubbard parameters, extended dynamical mean-field theory
(EDMFT)~\cite{smith2000,sun2002,nilsson2017} includes dynamical non-local
correlation into the self-consistent solution of the auxiliary problem.
Therefore, in EDMFT not only the local self-energy and Green's function are
updated in the self-consistency cycle, but also the local polarizability and
screened Coulomb potential. In EDMFT+$GW$, the self energy is expressed
as~\cite{nilsson2017}
\begin{equation}
    \Sigma^{\mathrm{EDMFT+\mathit{GW}}}_{kl} 
    = \Sigma^{\mathrm{EDMFT}}_{kl} + \Sigma^{\mathit{GW}}_{kl} - \Sigma^{\mathrm{dc}}_{kl}.
\end{equation}
According to Ref.\cite{nilsson2017}, the double counting is defined as
\begin{equation}\label{eq:si1}
    \Sigma^{\mathrm{dc}}_{kl}(\tau) = 
    -\sum_{mn}^{A} G_{mn}(\tau) W_{mknl}(\tau) =  -f^{A} G f^{A} W. 
\end{equation}
The self energy is then given as
\begin{equation}
    \begin{aligned}
    \Sigma_{kl}^{\mathrm{EDMFT+\mathit{GW}}} &= \Sigma^{\mathrm{EDMFT}} 
    -\sum^{A+E}_{mn} G_{mn}(\tau)W_{mknl}(\tau) + \sum^{A}_{mn} G_{mn}(\tau)W_{mknl}(\tau)\\
    &= \Sigma^{\mathrm{EDMFT}}_{kl} - \sum_{mn}^{E} G_{mn}(\tau)W_{mknl}(\tau).
    \end{aligned}
\end{equation}
Thus, we find that the double counting corrections adopted in EDMFT+$GW$ and
QDET are the same if the $G_0W_0$ approximation is employed for the environment
in EDMFT+$GW$. In that case, Eq.~\ref{eq:si1} becomes identical to Eq. 29 in the
main text.

\clearpage

\section{Implementation details}
\subsection{Implementation of exchange and correlation self-energy}\label{sec:sigmaxc}
The screened Coulomb potential $W_0$ is defined in Eq.~19 as
$W_0^{-1}= v^{-1} - P_0$, where $P_0$ is the irreducible polarizability.
Equivalently, it can be expressed in terms of the reducible polarizability $\chi
= P_0 + P_0 v \chi$ as $W_0 = v + v \chi v \equiv v +
W^{\mathrm{p}}$~\cite{govoni2015}. This separation of the screened Coulomb
interaction into $v$ and $W^{\mathrm{p}}(\omega)$ allows us to separate the
exchange-correlation self-energy $\Sigma_{\mathrm{xc}}$ in Eq. 17 of the main
text into the exchange part $\Sigma_{\mathrm{x}}$ and correlation part
$\Sigma_{\mathrm{c}}$. The former is given by
\begin{equation}
    { \left[ \Sigma_{\mathrm{x}} \right] }_{mn} = 
    - \sum_{k}^{A+E} n_k \braket{ \psi_m^{\mathrm{KS}} \psi_k^{\mathrm{KS}} 
    | v | \psi_n^{\mathrm{KS}} \psi_k^{\mathrm{KS}}  }  \label{eq:Sigma_x} ,
\end{equation}
where $\psi^{\mathrm{KS}}$ are the Kohn-Sham eigenstates and $n_k$ is the
occupation number of the state $\psi^{\mathrm{KS}}_k$. The correlation
contribution $\Sigma_{\mathrm{c}}$ is determined by evaluating Eq.~17 in the
main text with $W^{\mathrm{p}}$ instead of $W$. A compact expression for
$W^{\mathrm{p}}$ is obtained from the spectral decomposition of the dielectric
screening using its eigenvectors that we call the projective dielectric
eigenpotentials(PDEP)~\cite{wilson_efficient_2008,govoni2015}. The PDEPs form a
basis set $\{\ket{\varphi_i},i=1\cdots N_{\mathrm{PDEP}}\}$, where
$N_{\mathrm{PDEP}}$ is the number of basis functions, which allows us to write
$W^{\mathrm{p}}$ as
\begin{align}
\begin{split}
    W^{\mathrm{p}}(\omega) \approx 
    \Xi^{\mathrm{p}}(\omega) + \sum_{i,j=1}^{N_{\mathrm{PDEP}}} \ket{\tilde{\varphi_i}} \Lambda_{ij}(\omega) \bra{\tilde{\varphi_j}}.
\end{split}
\end{align}
where $\Xi^{\mathrm{p}}(\omega) = 4\pi e^2 \int_{R_{q=0}} \frac{dq}{{(2\pi)}^3}
\frac{\bar{\chi}_{00}}{q^2}$ is the head of $W^{\mathrm{p}}$, and
$\bar{\chi}_{00}$ is the head of the symmetrized polarizability
$\bar{\chi}=v^{1/2}\chi v^{1/2}$. The functions $\ket{\tilde{\varphi_i}} =
v^{\frac{1}{2}}\ket{\varphi_i}$ are the symmetrized PDEP basis functions. The
matrix elements $\Lambda_ij$ are elements of the body of $\bar{\chi}$. The
compact expression for $W^{\mathrm{p}}$ in the PDEP basis makes it possible to
calculate the Green's function and polarizability without explicit summations
over unoccupied states through the Lanczos algorithm~\cite{govoni2015}. 

Using contour deformation techniques, $\Sigma_{\mathrm{c}}$ is decomposed into
an integral $I$ along the imaginary axis and a residue part $R$.
The former can be further decomposed into three parts: $I_1$
from the head of $W_0$, $I_2$ from the body of $W_0$ with an explicit summation
over electronic states, and $I_3$ from the body of $W_0$ evaluated by the
Lanczos algorithm with $N_{\mathrm{Lanczos}}$ Lanczos chains. The residue part
$R$ is split into two parts: $R_1$ from the head of $W_0$ and $R_2$ from the
body of $W_0$. Thus, the correlation part of the self-energy is given by
\begin{align}
    { \left[ \Sigma_{\mathrm{c}} \right] }_{mn}(\omega) & = [I_1]_{mn}(\omega) + [I_2]_{mn}(\omega) + [I_3]_{mn}(\omega) +
[R_1]_{mn}(\omega) + [R_2]_{mn}(\omega),
\end{align}
where
\begin{align}
    & [I_1]_{mn}(\omega) = \delta_{mn}
    \int_{0}^{+\infty} 
    \frac{d\omega'}{\pi} \Xi^{\mathrm{p}}(\omega') 
    \left( \frac{ \left( \epsilon_n^{\mathrm{KS}} - \omega \right)}{ {\left( \epsilon_n^{\mathrm{KS}} - \omega \right)}^2 + {\omega'}^2 } \right) \\
    & [I_2]_{mn}(\omega) = \int_{0}^{+\infty} 
    \frac{d\omega'}{\Omega \pi}   
     \left( \sum_{k}^{N_{\mathrm{occ}}+N_{\mathrm{unocc}}} \sum_{i,j=1}^{N_{\mathrm{PDEP}}}
    \Lambda_{ij}(\omega') \braket{\varphi_m^i|\psi_k^{\mathrm{KS}}} \braket{\psi_k^{\mathrm{KS}}|\varphi_n^j} 
    \frac{\left( \epsilon_k^{\mathrm{KS}} - \omega \right)}{ {\left( \epsilon_k^{\mathrm{KS}} - \omega \right)}^2 + {\omega'}^2 } \right)
    \\
    & [I_3]_{mn}(\omega) = \int_{0}^{+\infty} 
    \frac{d\omega'}{\Omega \pi} \left(\sum_{i,j=1}^{N_{\mathrm{PDEP}}} \sum_{q_1,q_2=1}^{N_{\mathrm{Lanczos}}} \Lambda_{ij}(\omega') \braket{\varphi_m^i|{[\xi_n^j]}_{q_1}} {[U_n^j]}_{q_1 q_2} {[U_n^j]}_{1 q_2} 
    \frac{\left( {[\epsilon_n^j]}_{q_2} - \omega \right)}{ {\left( {[\epsilon_n^j]}_{q_2} - \omega \right)}^2 + {\omega'}^2 } \right) \\
    & [R_1]_{mn}(\omega) = \sum_{k}^{N_{\mathrm{occ}}+N_{\mathrm{unocc}}} \delta_{mn}  \delta_{nk}  f_k \Xi^{\mathrm{p}} (\epsilon_k^{\mathrm{KS}} - \omega) \\
    & [R_2]_{mn}(\omega) = \frac{1}{\Omega} \left(\sum_{k}^{N_{\mathrm{occ}}+N_{\mathrm{unocc}}} \sum_{i,j=1}^{N_{\mathrm{PDEP}}} 
    \Lambda_{ij}(\omega')  \braket{\varphi_m^i|\psi_k^{\mathrm{KS}}} f_k \braket{\psi_k^{\mathrm{KS}}|\varphi_n^j} \right) \\
& f_k = 
    \begin{cases}
    1-n_k \quad & \epsilon^{\mathrm{KS}}_{\mathrm{F}'} < \epsilon_k^{\mathrm{KS}} < \omega  \\
    -n_k \quad & \omega < \epsilon_k^{\mathrm{KS}} < \epsilon^{\mathrm{KS}}_{\mathrm{F}} \\
    0 \quad & \text{otherwise} .
    \end{cases}
\end{align}
Here $\ket{\xi_n^j}$, $\epsilon_n^j$ and $U_n^j$ are the Lanczos chain,
eigenvalue and unitary matrix for the right vector
$\ket{\varphi_n^j}=\ket{\psi_n^{\mathrm{KS}} \tilde{\varphi}_j^*}$,
respectively, using the Lanczos algorithm as presented in
Sec. 2.4 of Ref.~\citenum{govoni2015}. The quantities
$\epsilon_{\mathrm{F}}^{\mathrm{KS}}$ and $\epsilon_{\mathrm{F}'}^{\mathrm{KS}}$
are the energies of the highest partially occupied and highest fully occupied KS
orbital, respectively, $\Omega$ is the volume of the cell. 

\subsection{Implementation of double counting contribution to the self-energy}
We evaluate $\Sigma^{\mathrm{dc}}_{\mathrm{xc}}$ and
$\Delta\Sigma_{\mathrm{xc}}$ defined in Eq. 24 and 28 of
the main text, respectively, by introducing in Eq.~28 in the main text the
following expressions for $W$ and $G$: $W=v+W^\text{p}$,  and
$G_0=G_0^A+G_0^R$. This leads to the following definitions
\begin{align}
    & \Sigma^{\mathrm{dc}}_{\mathrm{xc}}(\omega) = \Sigma^{\mathrm{x,dc}} + \Sigma^{\mathrm{c,dc}}(\omega) \approx i \int_{-\infty}^{+\infty} 
    \frac{d\omega'}{2\pi} G_0^A(\omega+\omega')  v 
    + 
    i \int_{-\infty}^{+\infty} 
    \frac{d\omega'}{2\pi} G_0^A(\omega+\omega')  W^{\mathrm{p}}(\omega') \\
    & \Delta \Sigma_{\mathrm{xc}}(\omega) = \Delta \Sigma_{\mathrm{x}} + \Delta \Sigma_{\mathrm{c}}(\omega) \approx i \int_{-\infty}^{+\infty} 
    \frac{d\omega'}{2\pi} G_0^R (\omega+\omega')  v 
    + 
    i \int_{-\infty}^{+\infty} 
    \frac{d\omega'}{2\pi} G_0^R (\omega+\omega')  W^{\mathrm{p}}(\omega') .
\end{align}
The active space is formed by $N_{A}$ orbitals, while the
environment consists of $N_E$ explicit orbitals and the continuum of unoccupied
orbitals. As such, $N_{\mathrm{occ}}+N_{\mathrm{unocc}}=N_A+N_E$. Similar
derivations to those for $\Sigma_{\mathrm{xc}}$ (Sec.~\ref{sec:sigmaxc}) lead to
\begin{align}
    & {\left[ \Sigma^{\mathrm{dc}}_{\mathrm{x}} \right] }_{mn} = - \sum_{k}^{N_A} n_k \braket{ \psi_m^{\mathrm{KS}} \psi_k^{\mathrm{KS}} | v | \psi_n^{\mathrm{KS}} \psi_k^{\mathrm{KS}}  } \\
    & {\left[ \Sigma^{\mathrm{dc}}_{\mathrm{c}} \right] }_{mn}(\omega) = {[I_1]}_{mn}(\omega) +
    {[I^{\mathrm{dc}}_2]}_{mn}(\omega) + {[R_1]}_{mn}(\omega) + {[R^{\mathrm{dc}}_2]}_{mn}(\omega) \\
    & { \left[ \Delta \Sigma_{\mathrm{x}} \right] }_{mn} = - \sum_{k}^{N_E} n_k \braket{ \psi_m^{\mathrm{KS}} \psi_k^{\mathrm{KS}} | v | \psi_n^{\mathrm{KS}} \psi_k^{\mathrm{KS}}  } \\
    & { \left[ \Delta \Sigma_{\mathrm{c}} \right] }_{mn}(\omega) = {[\Delta I_2]}_{mn}(\omega) +
    {[I_3]}_{mn}(\omega) + {[\Delta R_2]}_{mn}(\omega) ,
\end{align}
where
\begin{align}
    & {[I^{\mathrm{dc}}_2]}_{mn}(\omega) = \int_{0}^{+\infty} 
    \frac{d\omega'}{\Omega\pi}   
     \left( \sum_{k}^{N_A} \sum_{i,j=1}^{N_{\mathrm{PDEP}}}
    \Lambda_{ij}(\omega') \braket{\varphi_m^i|\psi_k^{\mathrm{KS}}} \braket{\psi_k^{\mathrm{KS}}|\varphi_n^j} 
    \frac{\left( \epsilon_k^{\mathrm{KS}} - \omega \right)}{ {\left( \epsilon_k^{\mathrm{KS}} - \omega \right)}^2 + {\omega'}^2 } \right) \\
    & [R^{\mathrm{dc}}_2]_{mn}(\omega) = \frac{1}{\Omega} \left(\sum_{k}^{N_A} \sum_{i,j=1}^{N_{\mathrm{PDEP}}} 
    \Lambda_{ij}(\omega') f_k \braket{\varphi_m^i|\psi_k^{\mathrm{KS}}} \braket{\psi_k^{\mathrm{KS}}|\varphi_n^j} \right) \\
    & {[\Delta I_2]}_{mn}(\omega) = \int_{0}^{+\infty} 
    \frac{d\omega'}{\Omega\pi}   
     \left( \sum_{k}^{N_E} \sum_{i,j=1}^{N_{\mathrm{PDEP}}}
    \Lambda_{ij}(\omega') \braket{\varphi_m^i|\psi_k^{\mathrm{KS}}} \braket{\psi_k^{\mathrm{KS}}|\varphi_n^j} 
    \frac{\left( \epsilon_k^{\mathrm{KS}} - \omega \right)}{ {\left( \epsilon_k^{\mathrm{KS}} - \omega \right)}^2 + {\omega'}^2 } \right) \\
    & {[\Delta R_2]}_{mn}(\omega) = \frac{1}{\Omega} \left( \sum_{k}^{N_E} \sum_{i,j=1}^{N_{\mathrm{PDEP}}} 
    \Lambda_{ij}(\omega') f_k  \braket{\varphi_m^i|\psi_k^{\mathrm{KS}}} \braket{\psi_k^{\mathrm{KS}}|\varphi_n^j} \right) .
\end{align}

\clearpage

\section{Convergence of QDET calculations} \label{sec:convergence}

\begin{figure}[h]
\includegraphics{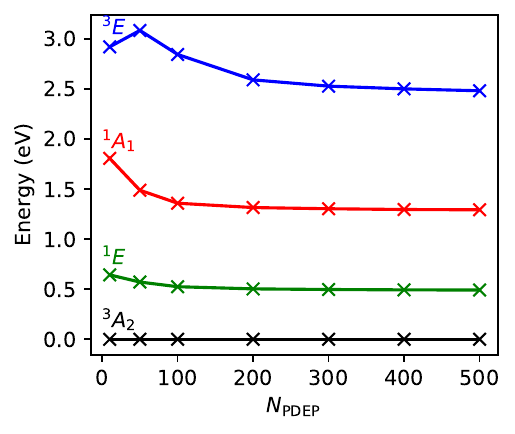}
\caption{\label{fig:npdep}Computed vertical excitation
  energies of the \ce{NV-} center in diamond as a function of the number of
  eigenpotentials $N_{\mathrm{PDEP}}$. Calculations are performed in a 511-atom
  cell using the PBE functional.}
\end{figure}

\subsection{Number of dielectric eigenpotentials} 
In Fig. \ref{fig:npdep} we show the convergence of the vertical excitation
energies for \ce{NV-} in a 511-atom cell using PBE with respect to the number of
PDEP eigenpotentials, $N_{\mathrm{PDEP}}$. A 6-orbital model (see Sec.
4 of the main text) is adopted. The plot shows that 500 PDEP
eigenpotentials yield converged excitation energies within 0.02 eV.

\subsection{Composition of the active space} 
As shown in Fig. (6) of the main text, localized defect
orbitals of the $\mathrm{SiV^0}$ defect in diamond are found in the valence
band, the band gap, and the conduction band of the host material. Ideally, all
localized orbitals should be included in the active space. However, the
computational complexity of full configuration interaction (FCI) grows
exponentially when increasing the number of occupied and unoccupied orbitals
simultaneously, making a straightforward convergence relative to all
localized orbitals impractical. To reduce
the computational cost, we do not include in the active space those localized
orbitals with energies above the conduction band minimum (CBM)) in all
calculations in the main text. To verify that excluding those localized orbitals
from the active space does not influence the results, we compare excitation
energies obtained with the minimal model~\cite{Thiering2019} with those of a
model including the most localized orbital within the conduction band. We find
that the exitation energies change by less than 0.05 eV. Hence, in all
calculations presented in the main text we only considered the
defect orbitals within the valence band, in addition to those in the minimal
model.

\clearpage

\section{Ghost states} \label{sec:GS}
We find that in QDET, unphysical excitations occur in some cases when conduction
and valence orbitals are included in the active space.  These unphysical
excitations, which we denote as \textit{ghost states}, correspond to excitations
from the defect to the conduction band, but with excitation energies that are
unphysically low. Below, for the example of the \ce{NV-} defect in diamond, we
analyze the origin of these ghost states. We note that our detailed analysis
reveals that ghost states occur with HFDC as well, and we will discuss those for
the example of the \ce{SiV^0} defect in diamond.

\subsection{\texorpdfstring{\ce{NV-}}{} in diamond}
\begin{figure}[t]
\includegraphics[width=1\textwidth]{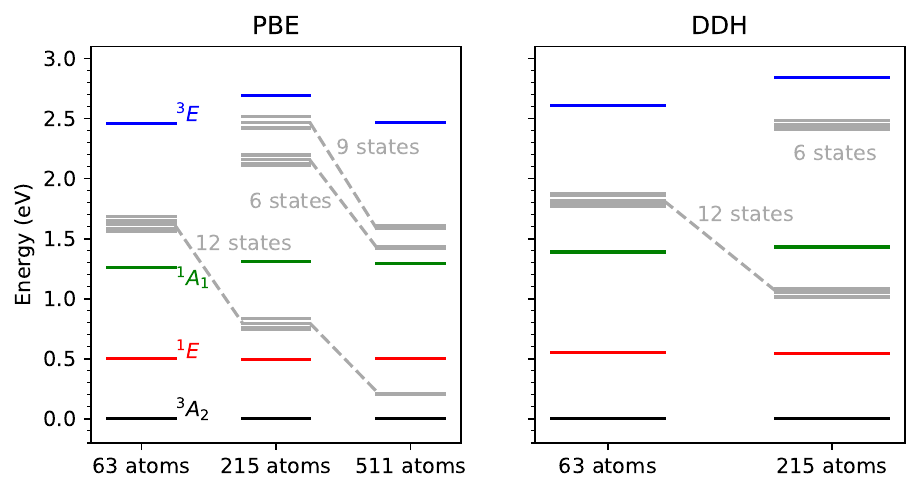}
\caption{\label{fig:ghost_supercell} 
Computed excitation energies (eV) for the \ce{NV-} center in diamond obtained
  from calculations with the PBE (left) and DDH functional (right). Ghost states
  (see text) are shown in grey. We find 12 nearly-degenerate single excitations
  from the $e$ states to the conduction band, 6 single excitations from the
  $a_1$ state to the conduction band, and 9 double excitations from the $e$
  state to the conduction band. The convergence of the ghost states is indicated
  with the dotted lines. }
\end{figure}
\begin{figure}[t!]
\includegraphics[width=1\textwidth]{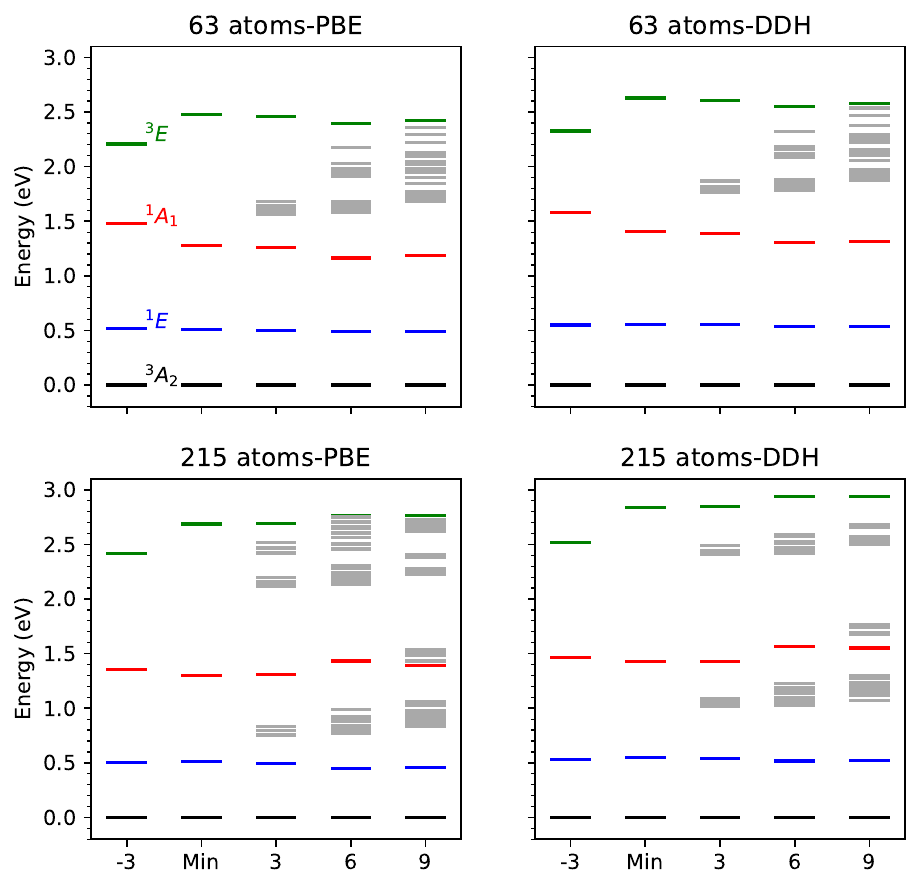}
\caption{\label{fig:ghost_active_space_pbe}
Excitation energies (eV) of the \ce{NV-} in diamond as a
  function of the active space size for a 63- (top) and 215-atom supercell
  (bottom). Calculations are performed with the PBE (left) and DDH functional
  (right). ``Min'' denotes the minimal model, ``-3'' indicates that three
  additional valance band orbitals are added to the minimal model, ``3''
  indicates that three additional conduction band orbitals are added to the
  minimal model, and so forth.
}
\end{figure}

Using \ce{NV-} as an example, we investigate the origin of ghost states. For the
63-atom and 215-atom supercells, we adopt a minimum model,
\textit{i.e.} we include in the active space the two defect $e$ and two defect
$a_1$ orbitals~\cite{ma2020}, and the three lowest conduction band orbitals. For
the 511-atom cell we adopt a model which includes 6 orbitals, as mentioned in
Sec. 4 of the main text, and the three lowest conduction
orbitals.

In Fig. \ref{fig:ghost_supercell} and \ref{fig:ghost_active_space_pbe}, we show
both the defect excitations and the ghost states as a function of the supercell
size and the active-space size, respectively. The energy of the ghost states
decreases with increasing supercell size but gradually increases with increasing
size of the active space. Considering that QDET is exact in the limit of an
infinitely large supercell and an infinitely large active space, the results
suggest that the slow convergence w.r.t. the active space size is the reason for
the existence of ghost states. Furthermore, we speculate that this problem may
be material dependent. As the slow convergence only affects the transitions to
the conduction-band orbitals, and not those within the defect, it us reasonable
to expect that the slow convergence originates from an inadequate description of
conduction-band orbitals when the Brillouin zone is only sampled at the
$\Gamma$-point. Diamond is an indirect bandgap semiconductor, and its conduction
band minimum requires an accurate $\mathbf{k}$-point sampling. 

\subsection{\texorpdfstring{\ce{V_S^0}}{} in ZnS}
\begin{figure}[t!]
\includegraphics{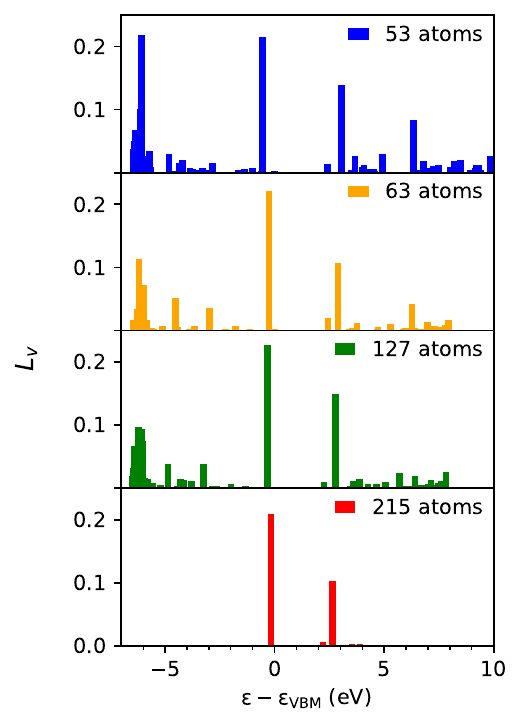}
\caption{\label{fig:vs0_localization_pbe} 
  Localization factor ($L_V$, see Eq. 30 in the main text)
  for the \ce{V_S^0} in ZnS as a function of the Kohn-Sham energy relative to
  the VBM. Results are shown for 4 different supercell sizes.}
\end{figure}
In order to test our hypothesis, we study the neutral
sulphur vacancy in cubic \ce{ZnS} (\ce{V_S^0}). As a direct bandgap material,
cubic \ce{ZnS} has a three-fold degenerate valence-band minimum (VBM) and a
non-degenerate conduction-band minimum (CBM). Again, we use the localization
factor defined in Eq.~30 of the main text and shown in Fig.
\ref{fig:vs0_localization_pbe} to determine the orbitals to
include in the active space. We find that a neutral sulphur vacancy introduces
four defect orbitals originating from the four dangling bonds of \ce{Zn}, one of
which is slightly lower than the VBM and a three-fold degenerate one slightly
above the CBM. We include the VBM and CBM in the active space to form an
eight-orbital active space. 

\begin{figure}[t!]
\includegraphics{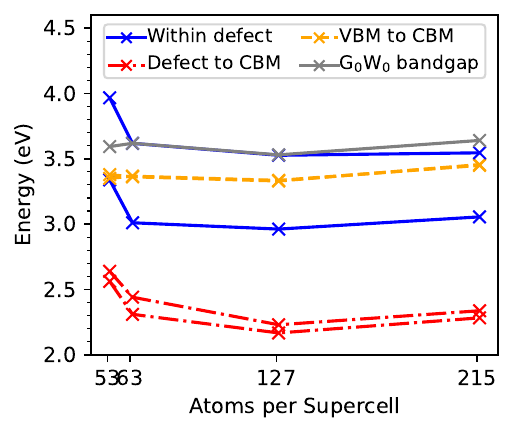}
\caption{\label{fig:vs0_supercell_pbe} Vertical excitation energies (eV) of the
  \ce{V_S^0} defect in Zns as functions of supercell size. }
\end{figure}

In Fig. \ref{fig:vs0_supercell_pbe} we show the vertical excitation energies of
\ce{V_S^0} as functions of supercell size. Similar to \ce{NV-}, the excitation
energies within the defect are relatively stable when increasing the supercell
size. In contrast to \ce{NV-}, both the excitation energies from the defect to
the CBM and those from the VBM to the CBM are also stable when increasing the
supercell size. This confirms our hypothesis that the unphysical excitations
found in diamond are due to an inaccurate description of the CBM due to the
indirect bandgap of the system.

In summary, we find excitations with
unphysically low energies occuring in QDET calculations for materials with
indirect bandgaps for supercells for which the $\Gamma$-point sampling is
insufficient to correctly sample the CBM. Our analysis shows that the
photoionization energies (excitation energies to the CBM) increase with
increasing active space size, but the convergence is slow. We find that these
energies are due to an inaccurate sampling of the CBM of an indirect-bandgap
material in the supercells used here. This finding is confirmed by our
investigation of the photoionization in the \ce{V_S^0} in the direct-bandgap
insulator ZnS, where we do not observe any ghost states.

\subsection{\texorpdfstring{\ce{SiV^0}}{} in diamond} \label{sec:spectrum}
\begin{figure}[t!]
\includegraphics{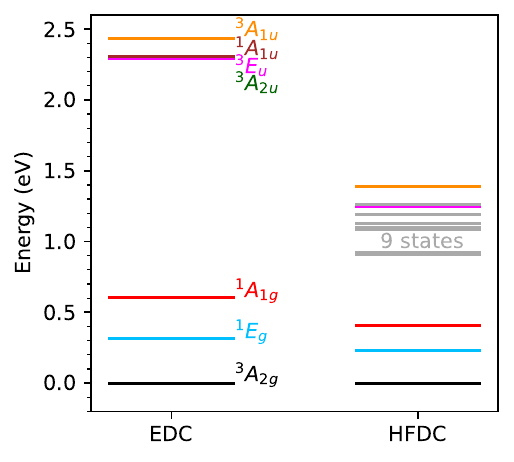}
\caption{\label{fig:ghost_dc_pbe_216} 
Vertical excitation energies for the \ce{SiV0} center in
  diamond in a 215-atom cell. Calculations are performed with the EDC (left) and
  HFDC (right) correction schemes (see text). The 9 ghost states in the
  calculations with HFDC are shown in grey. These ghost states represent
  excitations from the valence band to the defect orbitals.}
\end{figure}
Our analysis of ghost states in diamond reveals that ghost states occur in QDET
calculations using HFDC. In Fig. \ref{fig:ghost_dc_pbe_216} we present a
detailed comparison between the vertical excitation spectrum obtained when using
the two double counting schemes for \ce{SiV^0} in a 215-atom cell, using the
same computational setup as that of Ref. \citenum{ma2020}. Notably, HFDC
introduces 9 \textit{ghost states}, which correspond to excitations from the
valance band to the defect orbitals but have lower excitation energies than
those within the defect. We note that these \textit{ghost states} do not occur
in calculations with the EDC scheme. As the VBM of diamond is found at the
$\Gamma$-point in reciprocal space, these ghost states do not originate from the
bandstructure folding, but rather from the incomplete double counting correction
within HF. As the approximate HFDC scheme affects localized
defect orbitals differently than delocalized valence orbitals, excitations with
unphysical energies occur.

\clearpage

\section{Vertical excitation energies} \label{sec:table}

\begin{table*}[!h]
\caption{Vertical excitation energies (eV) of \ce{NV-}, \ce{SiV^0}, \ce{GeV^0},
  \ce{SnV^0} and \ce{PbV^0} in diamond. All calculations use the PBE functional.
  Experimental measurements of zero phonon line (ZPL) energies are shown in
  brackets in the last column. Reference  vertical  excitation  energies are
  computed from experimental ZPL when Stokes energies are available. Results are
  shown for calculations with the exact double counting (EDC) and Hartree-Fock
  double counting (HFDC) schemes (see text). Energy levels are labelled using
  the representations of the $C_{3v}$ point group.}
\begin{tabular}{ll|lll|lll|l}
\hline
System & Excitation & \multicolumn{3}{l}{EDC} & \multicolumn{3}{l|}{HFDC} & Ref \\
&  & \multicolumn{3}{l}{\# of atoms} & \multicolumn{3}{l|}{\# of atoms} & \\
 &  & 63 & 215 & 511 & 63 & 215 & 511 \\
\hline
\ce{NV-} 
    & ${}^1E_{}$ &  0.512 &      0.459 &  0.463 & 0.419 & 0.380 & 0.375 &                   \\
    & ${}^1A_{1}$ &  1.526 &      1.305 &  1.270 & 1.253 & 1.174 & 1.150 &                  \\
    & ${}^3E_{}$ &  1.944 &  2.023     &  2.152  & 1.516 & 1.381 & 1.324 &  2.180 \cite{Davies1976} (1.945 \cite{Davies1976}) \\
\hline
\ce{SiV^0} 
    & ${}^1E_{g}$ &   0.309 &      0.324 &   0.321 & 0.184 & 0.209 & 0.236 &                       \\
    & ${}^1A_{1g}$ &   0.604 &      0.645 &   0.642 & 0.326 & 0.380 & 0.435 &                      \\
    & ${}^3A_{2u}$ &   2.295 &      2.003 &   2.146 & 1.710 & 1.223 & 1.098 &                      \\
    & ${}^3E_{u}$ &   2.319 &      2.011 &   2.161 & 1.744 & 1.229 & 1.096 &     
    (1.31 \cite{Green2019})          \\
    & ${}^1A_{1u}$ &   2.365 &      2.040 &   2.183 & 1.798 & 1.251 & 1.111 &                      \\
    & ${}^3A_{1u}$ &   2.541 &      2.183 &   2.260 & 1.952 & 1.382 & 1.188 &                      \\
\hline
\ce{GeV^0} & ${}^1E_{g}$ &   0.363 &      0.334 &   0.357 & 0.243 & 0.260 & 0.289 &                     \\
    & ${}^1A_{1g}$ &   0.721 &      0.671 &   0.717 & 0.451 & 0.496 & 0.554 &                      \\
    & ${}^3A_{2u}$ &   3.062 &      2.661 &   2.924 & 2.098 & 1.579 & 1.456 &                      \\
    & ${}^3E_{u}$ &   3.069 &      2.653 &   2.925 & 2.117 & 1.573 & 1.443 &                       \\
    & ${}^1A_{1u}$ &   3.089 &      2.661 &   2.940 & 2.151 & 1.581 & 1.443 &                      \\
    & ${}^3A_{1u}$ &   3.235 &      2.756 &   2.970 & 2.281 & 1.676 & 1.495 &                      \\
\hline
\ce{SnV^0} & ${}^1E_{g}$ &   0.353 &      0.341 &   0.295 & 0.266 & 0.277 & 0.276 &                     \\
    & ${}^1A_{1g}$ &   0.711 &      0.686 &   0.596 & 0.510 & 0.538 & 0.551 &                      \\
    & ${}^3A_{2u}$ &   3.050 &      2.722 &   2.590 & 2.126 & 1.573 & 1.459 &                      \\
    & ${}^3E_{u}$ &   3.058 &      2.707 &   2.571 & 2.142 & 1.561 & 1.444 &         
                \\
    & ${}^1A_{1u}$ &   3.079 &      2.701 &   2.561 & 2.172 & 1.558 & 1.436 &                      \\
    & ${}^3A_{1u}$ &   3.192 &      2.796 &   2.616 & 2.281 & 1.654 & 1.491 &                      \\
\hline
\ce{PbV^0} & ${}^1E_{g}$ &   0.379 &      0.352 &   0.319 & 0.293 & 0.300 & 0.302 &                      \\
    & ${}^1A_{1g}$ &   0.751 &      0.704 &   0.640 & 0.563 & 0.587 & 0.600 &                      \\
    & ${}^3A_{2u}$ &   3.604 &      3.227 &   3.095 & 2.482 & 1.905 & 1.788 &                      \\
    & ${}^3E_{u}$ &   3.609 &      3.206 &   3.072 & 2.496 & 
    1.888 & 1.768 &                       \\
    & ${}^1A_{1u}$ &   3.623 &      3.193 &   3.056 & 2.520 & 1.879 & 1.755 &                      \\
    & ${}^3A_{1u}$ &   3.719 &      3.275 &   3.099 & 2.606 & 1.963 & 1.796 &                      \\
\hline
\end{tabular}
\label{tab:excitation_energies_full}
\end{table*}

\clearpage

\bibliography{ref}